\documentclass[lettersize,journal]{IEEEtran}
\usepackage{amsmath,amsfonts}
\usepackage[T1]{fontenc}
\usepackage{algorithmic}
\usepackage{setspace}
\usepackage{algorithm}
\usepackage{array}
\usepackage[caption=false,font=normalsize,labelfont=sf,textfont=sf]{subfig}
\usepackage{textcomp}
\usepackage{stfloats}
\usepackage{verbatim}
\usepackage{graphicx}
\usepackage{cite}
\usepackage[bottom]{footmisc}
\usepackage{balance}
\usepackage{amsthm}
\usepackage{mathtools}
\usepackage[utf8]{inputenc}
\usepackage{float}
\usepackage{ragged2e}
\usepackage[table,xcdraw]{xcolor}
\usepackage{multirow}
\usepackage{amssymb}

\newtheorem{lemma}{Lemma}
\newtheorem{remark}{Remark}
\newtheorem{corollary}{Corollary}
\DeclareMathOperator\erf{erf}

\begin{document}


\title{Optical RISs Improve the Secret Key Rate of Free-Space QKD in HAP-to-UAV Scenarios}

\author{Phuc V. Trinh,~\IEEEmembership{{\color{black}Senior} Member,~IEEE,} Shinya Sugiura,~\IEEEmembership{Senior Member,~IEEE,} 
Chao~Xu,~\IEEEmembership{Senior Member,~IEEE,} and Lajos~Hanzo,~\IEEEmembership{Life Fellow,~IEEE}
\vspace*{-7mm}
\thanks{P. V. Trinh and S.~Sugiura are with the Institute of Industrial Science, The University of Tokyo, Tokyo 153-8505, Japan (e-mail: \{trinh, sugiura\}@iis.u-tokyo.ac.jp). (\textit{Corresponding Author: Shinya Sugiura}.)}
\thanks{C. Xu and L. Hanzo are with the School of Electronics and Computer Science, University of Southampton, SO17 1BJ Southampton, U.K. (e-mail: \{cx1g08, lh\}@ecs.soton.ac.uk).}
\thanks{This work was supported in part by the JST ASPIRE (Grant JPMJAP2345), in part by the JSPS KAKENHI (Grant 23H00470, Grant 24K17272, Grant 24K21615), and in part by the Telecommunications Advancement Foundation (TAF).}
}



\maketitle

\begin{abstract}
Large optical reconfigurable intelligent surfaces (ORISs) are proposed for employment on building rooftops to facilitate free-space quantum key distribution (QKD) between high-altitude platforms (HAPs) and low-altitude platforms (LAPs). Due to practical constraints, the communication terminals can only be positioned beneath the LAPs, preventing direct upward links to HAPs. By deploying ORISs on rooftops to reflect the beam arriving from HAPs towards LAPs from below, reliable HAP-to-LAP links can be established. To accurately characterize the optical beam propagation, we develop an analytical channel model based on extended Huygens-Fresnel principles for representing both the atmospheric turbulence effects and the hovering fluctuations of LAPs. This model facilitates adaptive ORIS beam-width control through linear, quadratic, and focusing phase shifts, which are capable of effectively mitigating the detrimental effects of beam broadening and pointing errors (PE). {\color{black}Consequently, the information-theoretic bound of the secret key rate and the security performance of a decoy-state QKD protocol are analyzed}. Our findings demonstrate that quadratic phase shifts enhance the SKR at high HAP-ORIS zenith angles or mild PE conditions by narrowing the beam to optimal sizes. By contrast, linear phase shifts are advantageous at low HAP-ORIS zenith angles {\color{black}or} moderate-to-high PE by diverging the beam to mitigate LAP fluctuations.
\end{abstract}


\section{Introduction}
\IEEEPARstart{I}{n} parallel to the evolution of cellular systems from the fifth generation (5G) towards the sixth generation (6G), quantum information technology has also experienced rapid growth, particularly in quantum communications. The unconditional security offered by quantum key distribution (QKD) protocols is expected to substantially enhance the communication infrastructure of 6G networks \cite{Cao2022}. QKD utilizes quantum states to distribute cryptographic keys between legitimate parties, ensuring that any eavesdropping attempt perturbs the quantum states according to the physical laws of quantum mechanics, thereby revealing the eavesdropper's presence. QKD protocols are typically categorized as of discrete-variable (DV) and continuous-variable (CV) nature. More specifically, DV-QKD utilizes individual photons for encoding key information, relying on properties like polarizations and requiring single-photon sources as well as detectors \cite{Cao2022}. By contrast, CV-QKD maps the key to the quadrature components of Gaussian quantum states, necessitating coherent sources and homodyne/heterodyne detectors \cite{Hosseinidehaj2019,Liu2024}.

QKD systems have the capability of operating over both optical fiber and free-space optical (FSO) links, but FSO is capable of over-bridging thousands of kilometers from space to ground, eliminating the need for cable installations and relaying \cite{Trinh2018}. Briefly, fiber-based QKD faces more substantial pathloss in optical fibers than its FSO-based counterpart communicating over atmospheric channels \cite{Alberto2016}. Recent developments mark a significant shift in quantum network architecture, moving from terrestrial infrastructures to seamless integration with non-terrestrial platforms like unmanned aerial vehicles (UAVs) and satellites, forming essential parts of the \textit{Quantum Internet in the Sky} \cite{Trinh2024}. This paradigm shift aligns synergistically with the consideration of non-terrestrial networks (NTNs) in 6G, encompassing low-altitude platforms (LAPs), high-altitude platforms (HAPs), and low-Earth orbit (LEO) satellites{\color{black}\cite{Liu2018}. With the emergence of new connectivity paradigms between LAPs and HAPs \cite{Liu2018}, safeguarding these communication links with {\color{black}the aid of} QKD has become imperative. These links serve as critical intermediaries between terrestrial QKD systems and satellite-based quantum nodes, while extending secure coverage to remote regions {\color{black}with underdeveloped terrestrial} infrastructure. Shorter HAP-LAP distances further reduce {\color{black}the} link latency compared to satellites, while improving network resilience through redundant connections between HAPs and LAPs to guard against {\color{black}outages}.} While significant milestones have been reached in establishing quantum links from LEO satellites to the ground \cite{Lu2022}, research into establishing similar quantum links on other platforms, such as HAPs and LAPs, is in its infancy.

HAPs, relying either on fixed-wing or balloon-based aerial platforms operating at altitudes ranging from 19 km to 22 km in the stratosphere, have been designed for extended quasi-stationary flight {\color{black}(e.g., several months)}, powered by solar energy. HAPs can significantly expand the coverage of quantum networks, especially in challenging terrains{\color{black}\cite{Liu2018}}. On the other hand, LAPs rely on {\color{black}battery-powered} UAVs, such as rotary-wing drones, which operate at altitudes ranging from tens of meters to a few hundred meters, depending on national flight regulations. These platforms offer a flexible and agile infrastructure for quantum communications, making them ideal for rapid deployment in disaster zones and complex urban propagation environments. Recent progress has seen the theoretical exploration of HAPs-to-ground QKD links \cite{Chu2021}, and the experimental success of both drone-to-ground as well as of drone-to-drone QKD systems \cite{Liu2020,Liu2021}. However, the theoretical and experimental performance of HAPs-to-drone QKD links is unknown at the time of writing, leaving a connection gap in the global quantum NTNs.

While it would be desirable to install a communication terminal on top of a drone to establish a direct line-of-sight (LoS) link with HAPs, this is impractical. Explicitly, industrial drones are engineered to carry their payloads underneath using gimbals, which is convenient for optimizing communications with other terminals at similar altitudes or on the ground \cite{Liu2020,Liu2021}. The upper part of a drone houses critical components such as batteries, global positioning system (GPS) antennas, and the mechanical frame supporting the drone arms. Mounting equipment on top can interfere with the GPS signal and it is subject to significant vibrations and turbulence from the propellers. Additionally, placing high weight at the top may increase the risk of tipping over, especially in strong winds. By contrast, mounting equipment underneath the drone has the advantage of mitigating vibrations and allows LoS communications with the terminal on the ground. To create reliable QKD links with HAPs, innovative communication methods must be developed that allow the QKD terminal to be installed underneath the drone.
\vspace{-0.1cm}
\subsection{Related Studies}
\label{subsect:IA}
When the direct LoS links between a pair of optical transceivers are blocked or there are hardware alignment difficulties, optical reconfigurable intelligent surfaces (ORISs) \cite{Jamali2021,Wang2023} have been proposed for both classical \cite{Najafi2021,Ndjiongue2021,Wang2022,Wang2022-TVT,Wang2022-TVT-2,Chapala2022,Noh2023,Wang2023-JLT,Wang2023-TVT,Thang2023,Li2023,Ata2024,Ishida2024,Ajam2022,Sipani2023,Ajam2024} and quantum \cite{Kisseleff2023,Kundu2024} FSO scenarios. Compared to dedicated optical relay nodes, an ORIS offers a cost-effective alternative by using passive elements for controlling the phase of incident beams, hence enabling adaptive beam control and anomalous reflection in specific desired directions at low power consumption \cite{Minovich2015}. Thanks to its flat surface and compact electronics, an ORIS can be conveniently mounted on building walls or rooftops, typically relying on mirror-array and metasurface types \cite{Jamali2021,Wang2023}. Briefly, a mirror-array-based ORIS uses small mirrors on micro-electro-mechanical systems to control orientation, while a metasurface-based ORIS utilizes materials having optically modulated properties, like liquid crystals (LC), to produce phase shifts by modulating the molecular alignments.

To elaborate, previous studies typically design ORIS phase-shift profiles and model the optical beam propagation using geometric optics relying on far-field approximations \cite{Najafi2021,Ndjiongue2021,Wang2022,Wang2022-TVT,Wang2022-TVT-2,Chapala2022,Noh2023,Wang2023-JLT,Wang2023-TVT,Thang2023,Li2023,Ata2024,Ishida2024}. However, far-field approximations are only valid over distances of dozens of kilometers, which may not be suitable for practical ranges of intermediate-field FSO links. Fortunately, the Huygens-Fresnel (HF) principles have been exploited before for modeling ORIS-assisted FSO channels, which are valid for both intermediate and far fields, spanning distances from dozens of meters to dozens of kilometers \cite{Ajam2022}. Based on the classic HF principles, both linear phase shift (LPS) and quadratic phase shift (QPS) profiles across the ORIS were considered, where the LPS profile enables anomalous reflection of the beam according to the generalized Snell's law of reflection, while the QPS profile reduces beam divergence along the propagation path \cite{Ajam2022}. In designing the QPS profile, specific attention was dedicated both to pointing errors (PEs) arising from random misalignments at the transceivers and ORIS, as well as to beam non-orthogonality \cite{Sipani2023}. Most recently, a tractable power scaling law based on HF principles has been developed for LPS, QPS, and focusing phase shift (FPS) profiles, offering practical insights into the dependence of received power on the ORIS, on the receiving lens, and on beam widths \cite{Ajam2024}.

The concept of employing ORISs for enhancing non-LoS free-space terrestrial QKD systems was initially introduced in \cite{Kisseleff2023}. However, this proposal did not account either for the Gaussian power distribution of the optical beam or for the HF principles. In a recent development, the QPS profile, explicitly considering both the HF principles and the Gaussian power distribution of the optical beam, was investigated in free-space QKD {\color{black}non-LoS} terrestrial links \cite{Kundu2024}. Nevertheless, previous seminal studies have overlooked that both the geometric optics and the HF principles only characterize optical beam propagation in free space, i.e., in vacuum \cite{Ajam2022,Sipani2023,Ajam2024,Kundu2024}, but ignore the effects of atmospheric turbulence-induced beam broadening. Furthermore, while random misalignment-induced PEs were explored in \cite{Sipani2023} based on the HF principles for classical ORIS-aided FSO systems, the corresponding analysis of its QKD counterpart is unavailable in the literature.
\vspace{-0.2cm}
\subsection{Key Contributions}
\begin{table*}
\centering
\vspace{-0.3cm}
\captionsetup{font=footnotesize}
\caption{Comparison between this work and the state-of-the-art ORIS-aided classical FSO systems.}
\vspace{-0.1cm}
\scalebox{0.68}{\begin{tabular}{|c|ccc|cc|ccc|c|} 
\hline
\multirow{4}{*}{\textbf{Ref.}} &  \multicolumn{3}{c|}{\textbf{ORIS Design Principles}}  &     \multicolumn{2}{c|}{\begin{tabular}{@{}c@{}}\textbf{Beam Power}\\ \textbf{Distribution}\end{tabular}}  &  \multicolumn{3}{c|}{\begin{tabular}{@{}c@{}}\textbf{Pointing Errors}\\ \textbf{in 2 Orthogonal Axes}\end{tabular}} & \multirow{4}{*}{\begin{tabular}{@{}c@{}}\textbf{Non-}\\ \textbf{Terrestrial}\\ \textbf{Platforms}\end{tabular}} \\ \cline{2-9}
& \multicolumn{1}{c|}{\begin{tabular}{@{}c@{}}\textbf{Geometric optics}\\ (Far-field distances, \\vacuum channels)\end{tabular}} & \multicolumn{1}{c|}{\begin{tabular}{@{}c@{}}\textbf{HF principles}\\ (Intermediate \& far-field distances, \\vacuum channels)\end{tabular}} &\begin{tabular}{@{}c@{}}\textbf{Extended HF (EHF) principles}\\ (Intermediate \& far-field distances, \\atmospheric channels)\end{tabular}  & \multicolumn{1}{c|}{Uniform} &Gaussian  & \multicolumn{1}{c|}{\begin{tabular}{@{}c@{}}Uniform\end{tabular}} & \multicolumn{1}{c|}{\begin{tabular}{@{}c@{}}i.i.d.\\ Gaussian\end{tabular}} & \multicolumn{1}{c|}{\begin{tabular}{@{}c@{}}i.n.i.d.\\ Gaussian\end{tabular}} &  \\ \hline\hline
\cite{Najafi2021,Ndjiongue2021}& \multicolumn{1}{c|}{$\surd$} & \multicolumn{1}{c|}{} &  & \multicolumn{1}{c|}{} &$\surd$  &\multicolumn{1}{c|}{}  & \multicolumn{1}{c|}{$\surd$} & \multicolumn{1}{c|}{} &  \\ \hline
\cite{Wang2022} & \multicolumn{1}{c|}{$\surd$} & \multicolumn{1}{c|}{} &  & \multicolumn{1}{c|}{} &$\surd$  &\multicolumn{1}{c|}{}  &\multicolumn{1}{c|}{}  & \multicolumn{1}{c|}{} & \\ \hline
\cite{Wang2022-TVT,Wang2022-TVT-2,Chapala2022,Noh2023,Wang2023-JLT,Wang2023-TVT} & \multicolumn{1}{c|}{$\surd$} & \multicolumn{1}{c|}{} &  & \multicolumn{1}{c|}{} &$\surd$  &\multicolumn{1}{c|}{}  & \multicolumn{1}{c|}{$\surd$} & \multicolumn{1}{c|}{} & \\ \hline
\cite{Thang2023,Li2023,Ata2024} & \multicolumn{1}{c|}{$\surd$} & \multicolumn{1}{c|}{} &  & \multicolumn{1}{c|}{} &$\surd$  &\multicolumn{1}{c|}{}  & \multicolumn{1}{c|}{$\surd$} & \multicolumn{1}{c|}{} &$\surd$ \\ \hline
\cite{Ishida2024} & \multicolumn{1}{c|}{$\surd$} & \multicolumn{1}{c|}{} &  & \multicolumn{1}{c|}{$\surd$} &  &\multicolumn{1}{c|}{}  & \multicolumn{1}{c|}{$\surd$} & \multicolumn{1}{c|}{} & \\ \hline
\cite{Ajam2022} & \multicolumn{1}{c|}{} & \multicolumn{1}{c|}{$\surd$} &  & \multicolumn{1}{c|}{} &$\surd$  &\multicolumn{1}{c|}{}  &\multicolumn{1}{c|}{}   & \multicolumn{1}{c|}{} & \\ \hline
\cite{Sipani2023} & \multicolumn{1}{c|}{} & \multicolumn{1}{c|}{$\surd$} &  & \multicolumn{1}{c|}{} &$\surd $ &\multicolumn{1}{c|}{$\surd$}  & \multicolumn{1}{c|}{} & \multicolumn{1}{c|}{} & \\ \hline
\cite{Ajam2024} & \multicolumn{1}{c|}{} & \multicolumn{1}{c|}{$\surd$} &  & \multicolumn{1}{c|}{} &$\surd$  &\multicolumn{1}{c|}{}  &\multicolumn{1}{c|}{}   & \multicolumn{1}{c|}{} & \\ \hline\hline
This work & \multicolumn{1}{c|}{} & \multicolumn{1}{c|}{} &\multicolumn{1}{c|}{$\surd$}  & \multicolumn{1}{c|}{} &$\surd$  &\multicolumn{1}{c|}{}  & \multicolumn{1}{c|}{$\surd$} & \multicolumn{1}{c|}{$\surd$} &$\surd$ \\ \hline
\end{tabular}}
\label{Table1}
\vspace{-0.1cm}
\end{table*}
\begin{table*}[tp]
\begin{center}
\captionsetup{font=footnotesize}
\caption{Comparison between this work and the state-of-the-art ORIS-aided quantum FSO systems.}
\vspace{-0.3cm}
\scalebox{0.68}{ \begin{tabular}{|c|cccccc|cc|c|c|c|}
\hline
\multirow{3}{*}{\textbf{Ref.}} &  \multicolumn{6}{c|}{\textbf{ORIS Design Principles}} &  \multicolumn{2}{c|}{\begin{tabular}{@{}c@{}}\textbf{Beam Power}\\ \textbf{Distribution}\end{tabular}} &\multirow{3}{*}{\begin{tabular}{@{}c@{}}\textbf{Pointing}\\ \textbf{Errors}\end{tabular}}  &\multirow{3}{*}{\begin{tabular}{@{}c@{}}\textbf{Non-}\\\textbf{Terrestrial}\\ \textbf{Platforms}\end{tabular}}  & \multirow{3}{*}{{\color{black}\textbf{Finite-key effects}}}  \\ \cline{2-9}
&\multicolumn{3}{c|}{\textbf{HF Principles}} & \multicolumn{3}{c|}{\textbf{EHF Principles}} & \multicolumn{1}{c|}{\multirow{2}{*}{Uniform}} & \multirow{2}{*}{Gaussian}  &  &  &  \\ \cline{2-7}
& \multicolumn{1}{c|}{LPS} & \multicolumn{1}{c|}{QPS} & \multicolumn{1}{c|}{FPS} & \multicolumn{1}{c|}{LPS} & \multicolumn{1}{c|}{QPS} &FPS  & \multicolumn{1}{c|}{} &  &  &  & \\ \hline \hline
\cite{Kisseleff2023}& \multicolumn{1}{c|}{} & \multicolumn{1}{c|}{} & \multicolumn{1}{c|}{} & \multicolumn{1}{c|}{} & \multicolumn{1}{c|}{} &  & \multicolumn{1}{c|}{$\surd$} &  &  &  & \\ \hline
\cite{Kundu2024} & \multicolumn{1}{c|}{} & \multicolumn{1}{c|}{$\surd$} & \multicolumn{1}{c|}{} & \multicolumn{1}{c|}{} & \multicolumn{1}{c|}{} &  & \multicolumn{1}{c|}{} &$\surd$  &  &  & \\ \hline  \hline
This work& \multicolumn{1}{c|}{} & \multicolumn{1}{c|}{} & \multicolumn{1}{c|}{} & \multicolumn{1}{c|}{$\surd$} & \multicolumn{1}{c|}{$\surd$} &$\surd$  & \multicolumn{1}{c|}{} &$\surd$  &$\surd$  &$\surd$  &{\color{black}$\surd$}  \\ \hline
\end{tabular}}
\label{Table2}
\vspace{-0.5cm}
\end{center}
\end{table*}
In this paper, we propose, for the first time, the utilization of an ORIS for supporting QKD links between HAPs and LAPs. Specifically, an ORIS is strategically positioned on a building rooftop to reflect the signal impinging from HAPs towards LAPs from below. This configuration allows the communication terminal to be installed underneath the LAP. The rooftop positioning is more practical than wall mounting, as it provides unobstructed views and ample space{\footnote{Recent studies \cite{Thang2023,Li2023,Ata2024} reported on the deployment of ORIS on LAPs. However, this approach faces significant challenges. The limited space on the LAP confines ORISs to a size much smaller than the incoming beam, inevitably causing severe geometrical loss. Additionally, the substantial PEs imposed by the LAP's hovering fluctuations may result in high pathloss and frequent outages at the receiving end.}}. The LAP is typically an industrial rotary-wing drone capable of carrying substantial payloads \cite{Liu2020,Liu2021}. Additionally, ORIS enables adaptive beam-width control through adaptive phase-shift profiles, including LPS, QPS, and FPS. Tables \ref{Table1} and \ref{Table2} boldly contrast our work against the current state-of-the-art ORIS-aided FSO systems in classical and quantum communication scenarios, respectively. Our contributions in this paper can be summarized in more depth as follows.
\begin{itemize}
\item To accurately characterize optical beam propagation over atmospheric channels, we employ the extended HF (EHF) principles \cite{Andrews2005,Lutomirski1971,Ricklin2002} to model the effects of atmospheric turbulence-induced beam broadening and various phase-shift profiles on the received beam-width at the LAP. As an extension of the classic HF principles, the EHF principles are applicable to both intermediate-field and far-field distances, covering all practical FSO link ranges. Furthermore, the proposed EHF model accurately characterizes the Gaussian power intensity profile of the FSO beam incident upon the ORIS, which is fundamentally different from the uniform profile of radio-frequency RIS systems. This distinction is crucial for ORIS designs since both classical and quantum FSO systems use a coherent laser source having a Gaussian profile \cite{Ajam2022,Kundu2024}.
\item The hovering fluctuations of LAPS, caused by GPS inaccuracies or strong winds, lead to significant PEs in the ORIS-to-LAP link. Our analytical framework incorporates these fluctuations in two orthogonal axes by modeling them as two independent but not identically distributed (i.n.i.d.) Gaussian random variables (RVs)\footnote{This fact was validated by actual drone-based FSO experiments in \cite{Trinh2021}.}. We derive a closed-form expression for the statistical geometric and misalignment loss (GML), which is corroborated by Monte-Carlo (MC) simulations. Remarkably, previous studies typically assume simplified PE scenarios associated with independent and identically distributed (i.i.d.) Gaussian RVs in two orthogonal axes \cite{Najafi2021,Ndjiongue2021,Wang2022-TVT,Wang2022-TVT-2,Chapala2022,Noh2023,Wang2023-JLT,Wang2023-TVT,Thang2023,Li2023,Ata2024,Ishida2024}, or tractable uniformly distributed fluctuations \cite{Sipani2023}. Our model, therefore, provides a more generalized approach for analyzing the PE of ORIS-aided FSO systems.
\item Leveraging the newly developed framework based on EHF principles and generalized PEs, we formulate the ultimate information-theoretic bound of the secret key rate (SKR) in an HAP-ORIS-LAP QKD system. We examine the average Pirandola-Laurenza-Ottaviani-Banchi (PLOB) bound, representing the theoretical upper limit for the SKR of any QKD protocols, including both DV and CV systems. In contrast to previous studies that use a single atmospheric model for all transmission paths \cite{Ajam2022,Ajam2024,Kundu2024}, we consider independent atmospheric turbulence statistics for the HAP-ORIS and ORIS-LAP paths, giving cognizance to their distinct distances and atmospheric profiles. {\color{black}Notably, the total probability distribution of the channel transmittance is newly derived, incorporating random effects from both atmospheric turbulence and generalized PE. Under these conditions,} we extensively investigate all phase-shift profiles, including the LPS, QPS, and FPS, to optimize the received beam-width at the LAP, leading to improved SKR under various operational conditions. Thus, we provide a valuable framework for the engineering design of HAP-to-LAP QKD links.
\item {\color{black}To further highlight the potential application of the proposed theoretical framework, we provide a comprehensive analysis of the two-decoy-state DV-QKD protocol, incorporating finite-key effects that account for statistical uncertainties and reduced key rates due to the limited number of exchanged quantum signals. This analysis is particularly crucial for HAP-to-LAP links, where the battery-powered LAP has a constrained operational duration. In such scenarios, the absence of an arbitrarily large number of received signals makes it essential to address statistical uncertainties and carefully handle the statistical bounds of parameter fluctuations. Leveraging finite-key considerations, we present numerical results for the quantum bit error rate (QBER) and the secret key length achievable within the LAP's operational time. To the best of our knowledge, this treatise is the first one to provide such results for HAP-to-LAP QKD scenarios.}
\end{itemize}
The remainder of this paper is organized as follows. Section \ref{sect:Section2}  introduces the system and channel models of an ORIS-aided HAP-to-LAP QKD link, {\color{black}and examines the influence of ORIS on quantum signals during operational control.} In Section \ref{sect:Section3}, we develop the analytical framework for modeling the end-to-end GML based on the EHF principles, considering practical ORIS phase-shift profiles and generalized PEs at the LAP. In Section \ref{sect:Section4}, a novel closed-form expression of the PLOB bound is derived for the SKR metric, {\color{black}and a comprehensive security analysis of the two-decoy-state DV-QKD protocol incorporating finite-key effects is also presented.} Detailed numerical results and discussions are provided in Section \ref{sect:Section5}. Finally, Section \ref{sect:Section6} concludes the paper.

\textit{Notations:} Vectors and matrices are represented by boldface lowercase and uppercase letters, respectively; $\left | \cdot \right |$ denotes the absolute value, and $\left \| \mathbf{x} \right \|\!\!\!=\!\!\!\sqrt{x_{1}^{2}\!+\!x_{2}^{2}\!+\!\cdots \!+\!x_{n}^{2}}$ is the norm of a vector $\mathbf{x}\!\!\!=\!\!\!\left (x_{1},x_{2},\!\cdots\!,x_{n}  \right )$; $j$ denotes the imaginary unit and $\mathcal{R}\!\left \{ \cdot \right \}$ is the real part of a complex number; $\mathbb{E}[\cdot]$ denotes the statistical expectation; $\left \langle \cdot  \right \rangle$ represents the ensemble average. $x\!\!\sim\!\! \mathcal{N}(\mu_x,\sigma_x^2)$ indicates that the RV $x$ follows the Gaussian distribution having statistical mean $\mu_x$ and variance $\sigma_x^2$; $y\!\!\sim \!\!\mathcal{LN}(\mu_y,\sigma_y^2)$ means that the RV $y$ follows the log-normal distribution with statistical mean $\mu_y$ and variance $\sigma_y^2$. Finally, $\text{erf}\!\left ( z \right )\!\!=\!\!\frac{2}{\sqrt{\pi}}\!\int_{0}^{z}\!\exp\!\left ( -t^{2} \right )\!\text{d}t$ is the Gaussian error function, {\color{black}and $\text{erfc}\!\left ( z \right )\!=\!1-\text{erf}\!\left ( z \right )$ denotes the complementary error function.}
\vspace{-0.1cm}
\section{System and Channel Models}
\label{sect:Section2}
\subsection{System Model}
\begin{figure}[t]
\centering
\includegraphics[scale=0.35]{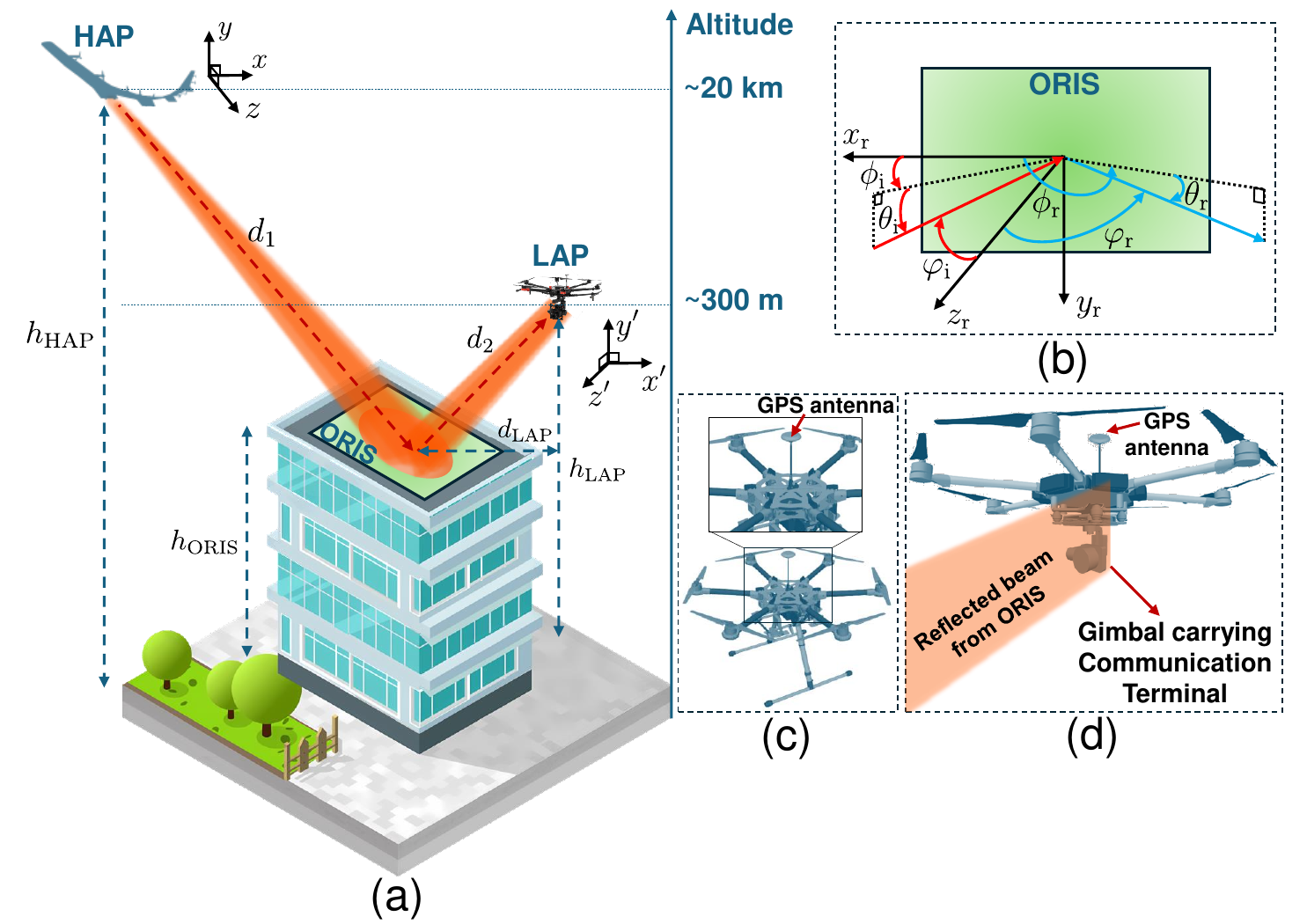}
\caption{(a) Schematic model of the ORIS-aided HAP-to-LAP QKD link with ORIS deployed on a building rooftop; (b) ORIS coordinates and beam propagation angles; (c) LAP's top view; (d) LAP's side view.}
\label{fig_1}
\vspace{-0.4cm}
\end{figure}

We investigate a quantum NTN downlink scenario, where a HAP seeks to establish a QKD link with a LAP, specifically with a rotary-wing drone. Due to practical constraints, the QKD terminal is mounted underneath the drone, optimizing communications with other terminals at similar or lower altitudes \cite{Liu2020,Liu2021}, but impeding signal reception from the HAP. To overcome this limitation, a large ORIS is placed on a building rooftop at a lower altitude than the drone, enabling the reflection of the incoming signal from the HAP towards the drone from below\footnote{This concept is also applicable to the uplink scenario, where a LAP, i.e., a drone, transmits signals to a HAP by directing the beam towards an ORIS, which then reflects the signals upwards. Deploying the ORIS on a building rooftop offers ample space for a large ORIS installation, avoiding the potential obstruction issues of a wall-mounted ORIS. The ORIS effectively reduces geometrical losses by fully reflecting the broadened optical beam. In this paper, we assume that the ORIS is large enough to capture the entire optical beam, with its specific dimensions detailed in Section \ref{sect:Section3a}.}, as depicted in Fig. \ref{fig_1}a. The transmitter (Tx) on the HAP is positioned at the origin of the $xyz$-coordinate system at an altitude of $h_{\text{HAP}}$, while the receiver (Rx) on the drone at an altitude of $h_{\text{LAP}}$ is located at the origin of the $x'y'z'$-coordinate system. The center of the ORIS, situated at an altitude of $h_{\text{ORIS}}$, is positioned at the origin of the $x_{\text{r}}y_{\text{r}}z_{\text{r}}$ coordinate system, as illustrated in Fig. \ref{fig_1}b, with the horizontal distance from the ORIS center to the projection of LAP on the $x_{\text{r}}y_{\text{r}}$-plane denoted as $d_{\text{LAP}}$. The $x_{\text{r}}y_{\text{r}}$-plane is parallel to the $xz$-plane and the $z_{\text{r}}$-axis is parallel to the $y$-axis. The Tx is equipped with a laser source that emits an optical beam having a Gaussian power density profile. The beam axis transmitted from Tx intersects the $x_{\text{r}}y_{\text{r}}$-plane of the ORIS at a distance $d_1$, and it is oriented in the direction $\mathbf{\Psi}_{\text{i}} \!=\! (\theta_{\text{i}}, \phi_{\text{i}})$, where $\theta_{\text{i}}$ represents the elevation angle between the $x_{\text{r}}y_{\text{r}}$-plane and the beam axis, while $\phi_{\text{i}}$ denotes the angle between the projection of the beam axis onto the $x_{\text{r}}y_{\text{r}}$-plane and the $x_{\text{r}}$-axis, as depicted in Fig. \ref{fig_1}b. In addition, we define $\varphi_{\text{i}}\!=\!90^{\circ}\!-\!{\color{black}\theta_{\text{i}}}$ as the zenith angle between the beam axis and the $z_{\text{r}}$-plane. The beam reflected from the ORIS towards the drone at a distance $d_2$ is directed towards $\mathbf{\Psi}_{\text{r}}\! =\! (\theta_{\text{r}}, \phi_{\text{r}})$, where $\theta_{\text{r}}$ is the angle between the $x_{\text{r}}y_{\text{r}}$-plane and the reflected beam axis. Furthermore, $\phi_{\text{r}}$ denotes the angle between the projection of the reflected beam axis onto the $x_{\text{r}}y_{\text{r}}$-plane and the $x_{\text{r}}$-axis. Similarly, we define $\varphi_{\text{r}}\!=\!90^{\circ}\!-\!{\color{black}\theta_{\text{r}}}$ as the zenith angle between the reflected beam axis and the $z_{\text{r}}$-plane.

Similar to \cite{Ajam2022,Ajam2024}, without loss of generality, we assume $\phi_{\text{i}}\!=\!0$ and $\phi_{\text{r}}\!=\!\pi$ for analytical tractability\footnote{The shapes of the beam incident upon the ORIS and the projection of the beam reflected to the Rx aperture on the ORIS are both ellipses, rotated by angles $\phi_{\text{i}}$ and $\phi_{\text{r}}$, respectively. When $\phi_{\text{i}}\!=\!0$ and $\phi_{\text{r}}\!=\!\pi$, the axes of both ellipses coincide, maximizing the received power \cite{Ajam2024}. Therefore, the results presented in this paper serve as upper bounds for the general case.}. Finally, we assume that the Rx communication terminal cannot be mounted on top of the drone due to the space limited by the GPS antennas and owing to the significant vibrations from the propellers (Fig. \ref{fig_1}c). Consequently, in practice, it can only be deployed on a 3-axis gimbal attached beneath the drone (Fig.~\ref{fig_1}d). This practical issue has often been overlooked in the literature. Recent advances in miniaturized optics and engineering have led to the successful development of real-world prototypes for compact communication terminals\footnote{These advanced miniaturized optical terminals feature a fine-pointing and tracking system based on the closed-loop operation of a position detector and a fast-steering mirror. This system corrects angle-of-arrival fluctuations caused by beam non-orthogonality and displacements at the Rx aperture, ensuring stable and accurate free-space beam coupling into a small detector or optical fiber \cite{Alberto2022,Kolev2023,Alberto2024}.}, which can be installed on various small platforms, such as nanosatellites, HAPs, and drones \cite{Alberto2022,Kolev2023,Alberto2024}. This confirms that FSO systems have become a reality for aerial platforms.

We consider an ORIS of size $\sum_{\text{ORIS}}\!\!=\!\!L_x \!\!\times \!\!L_y$, where $L_x$ and $L_y$ are the dimensions of the ORIS along the $x_{\text{r}}$- and $y_{\text{r}}$-axes, respectively. This ORIS is of the metasurface type, composed of LC molecules, which act as passive sub-wavelength elements designed to manipulate the properties of the incident beam. Given that typically we have $L_x, L_y \!\gg\! \lambda$, where $\lambda$ is the optical wavelength\footnote{In both classical and QKD systems, common optical wavelengths are 810 nm and 1550 nm \cite{Trinh2024}, which are considered eye-safe under the IEC 60825-1 standard, classified as class 1M \cite{Ghassemlooy2012}. In the HAP-to-drone scenario depicted in Fig. \ref{fig_1}a, the entire incoming optical beam is contained within the ORIS dimensions, which is then reflected skyward towards the drone. This setup ensures complete safety, as it prevents human exposure to the concentrated beam when QPS and FPS profiles are applied at the ORIS.}, a metasurface-based ORIS can be modeled as a continuous surface with continuous phase-shift profiles \cite{Ajam2024, Kundu2024}. For the ORIS designs, we consider the following phase-shift profiles.
\begin{itemize}
\item \textbf{\textit{LPS profile:}} This profile facilitates the generalized Snell's law of reflection and redirection of the beam from the Tx to the Rx by utilizing an ORIS phase-shift profile that varies linearly along the $x_{\text{r}}$- and $y_{\text{r}}$-axes as follows \cite{Ajam2022,Ajam2024}.
\begin{align}
\label{eq:LP}
\Phi _{\text{ORIS}}^{\text{LP}}\!\left ( \mathbf{r}_{\text{r}} \right )=k\left ( \Phi _{x}x_{\text{r}} + \Phi _{y}y_{\text{r}}+\Phi _{0} \right ),
\end{align}
where $k=2\pi/\lambda$ denotes the wave number and $ \mathbf{r}_{\text{r}} =\left(x_{\text{r}}, y_{\text{r}}, 0\right)$ represents a point in the $x_{\text{r}}y_{\text{r}}$-plane, and $\Phi _{0}$ is constant. The optical beam emerging from Tx in the direction $\mathbf{\Psi}_{\text{i}}$ is redirected to the Rx direction $\mathbf{\Psi}_{\text{r}}$ by applying the phase shift gradients as
\begin{align}
\Phi _{x}&=\cos\left ( \theta _{\text{i}} \right )\cos\left ( \phi_{\text{i}} \right )+\cos\left ( \theta _{\text{r}} \right )\cos\left ( \phi_{\text{r}} \right ),\\
\Phi _{y}&=\cos\left ( \theta _{\text{i}} \right )\sin\left ( \phi_{\text{i}} \right )+\cos\left ( \theta _{\text{r}} \right )\sin\left ( \phi_{\text{r}} \right ).
\end{align}
\item \textbf{\textit{QPS profile:}} This profile focuses the optical beam at a distance $f$ from the ORIS in the direction $\mathbf{\Psi}_{\text{r}}$, reducing the beam width of the reflected beam by applying a phase-shift profile that changes quadratically along the $x_{\text{r}}$- and $y_{\text{r}}$-axes as follows \cite{Ajam2022,Ajam2024}.
\begin{align}
\label{eq:QP}
\Phi _{\text{ORIS}}^{\text{QP}}\!\left ( \mathbf{r}_{\text{r}} \right )\!=\!k\!\left ( \Phi _{x^2}x_{\text{r}}^2 \!+\! \Phi _{y^2}y_{\text{r}}^2\!+\!\Phi _{x}x_{\text{r}} \!+\! \Phi _{y}y_{\text{r}}\!+\!\Phi _{0} \right )\!,
\end{align}
where the terms $\Phi _{x^2}$ and $\Phi _{y^2}$ are given by
\begin{align}
\Phi _{x^2}&=-\frac{\sin^2\left ( \theta _{\text{i}} \right )}{2R\left ( d_{1} \right )}-\frac{\sin^2\left ( \theta _{\text{r}} \right )}{2d_{2}}+\frac{\sin^2\left ( \theta _{\text{r}} \right )}{4f}\\
\Phi _{y^2}&=-\frac{1}{2R\left ( d_{1} \right )}-\frac{1}{2d_{2}}+\frac{1}{4f},
\end{align}
where $R\left ( d_{1} \right )$ is the radius of curvature at the distance $d_1$ along the HAP-ORIS path. The term $\frac{1}{4f}$ introduces a parabolic phase profile that narrows the beam at a focus distance $f$. Beyond this point, the beam becomes divergent.
\item \textbf{\textit{FPS profile:}} This profile focuses the optical beam at the Rx, functioning like an artificial lens to concentrate the incident beam at a distance of $d_2$ by utilizing the phase-shift profile that eliminates the accumulated phase of the incident beam as follows \cite{Ajam2022,Ajam2024}.
\begin{align}
\label{eq:FP}
\Phi _{\text{ORIS}}^{\text{FP}}=-\psi _{\text{in}}-k\left \| \tilde{\mathbf{r}}_{\text{o}} -\mathbf{r}_{\text{r}}\right \|,
\end{align}
where $\psi _{\text{in}}$ denotes the phase of the incident beam on the ORIS and $\tilde{\mathbf{r}}_{\text{o}}\!=\!\left ( \tilde{x}_{\text{o}}, \tilde{y}_{\text{o}}, \tilde{z}_{\text{o}} \right ) \!=\!\left ( -d_{2}\cos\left ( \theta_{\text{r}} \right ),0,d_{2}\sin\left ( \theta_{\text{r}} \right ) \right )$ is the center of the Rx aperture.
\end{itemize}
\vspace{-0.3cm}
\subsection{Channel Model}
In linear quantum optics, the losses can be characterized by the input-output relationship of \cite{Weedbrook2012}
\begin{align}
\label{eq:input_output}
\hat{a}_{\text{out}}=\sqrt{\tau }\hat{a}_{\text{in}}+\sqrt{1-\tau }\hat{c},
\end{align}
where $\hat{a}_{\text{in}}$ and $\hat{a}_{\text{out}}$ are the input and output field annihilation operators, respectively. Furthermore, $\hat{c}$ is the environmental mode operator in the vacuum state, and $\tau$ is the channel transmittance, characterizing the linear losses within the channel. From (\ref{eq:input_output}), $\tau$ is confined to the range of $\left [ 0,1 \right ]$ for preserving the canonical commutation relations for the quantized optical field operators in the input-output relationship. In free-space QKD systems, quantum signals are transmitted through atmospheric channels. Consequently, $\tau$ characterizes the fluctuating loss, which is treated as a random variable. The input-output relationship in (\ref{eq:input_output}) can be transformed into the Schrodinger picture of motion to derive the corresponding density operators \cite{Semenov2009}. By employing the Glauber-Sudarshan $P$ representation \cite{Glauber1963,Sudarshan1963}, the relationship between the quantum states transmitted and received through atmospheric channels {\color{black}is} described as \cite{Semenov2009}
$P_{\text{out}}\!\left ( \alpha  \right )\!\!=\!\!\int f\!\left ( \tau  \right )\!\frac{1}{\tau }\!P_{\text{in}}\!\left ( \frac{\alpha}{\sqrt{\tau }}  \right )\!\text{d}\tau$,
where $P_{\text{in}}\!\left ( \alpha  \right )$ and $P_{\text{out}}\!\left ( \alpha  \right )$ are the input and output $P$ functions, respectively. Furthermore, $f\!\left ( \tau  \right )$ denotes the probability distribution of transmittance (PDT). It is {\color{black}recognized} that characterizing quantum signals received from atmospheric channels reduces to the accurate modeling of the PDT. In this paper, the quantum atmospheric channel transmittance $\tau$ is assumed to represent four degradation factors, formulated as
\begin{align}
\label{eq:tau}
\tau =\tau _{\text{eff}}{\color{black}\tau_{\text{ORIS}}}\tau _{\text{l}}I_{\text{a}}\tau_{\text{p}},
\end{align}
where $\tau _{\text{eff}}$ is the Rx efficiency, {\color{black}$\tau_{\text{ORIS}}$ is the ORIS reflectance,} $\tau _{\text{l}}$ is the deterministic loss over the atmosphere, $I_{\text{a}}$ is the random intensity fluctuation due to atmospheric turbulence, and $\tau_{\text{p}}$ is the GML affected by the ORIS phase-shift profiles and drone hovering fluctuation-induced PE.

For elevation angle $\theta_{\text{i}}>20^{\circ}$, let $\tau _{\text{l,1}}$ denote the atmospheric loss in the HAP-ORIS slanted path, which is scaled as \cite{Dequal2021}
\begin{align}
\label{eq:loss1}
\tau _{\text{l,1}} =\tau _{\text{zen}}^{\sec\left ( \varphi _{\text{i}} \right )},
\end{align}
where $\tau_{\text{zen}}$ denotes the transmission efficiency at $\varphi _{\text{i}}\!=\!0^{\circ}$, which can be conveniently estimated by the popular MODTRAN code \cite{Dequal2021}, which is a widely used atmospheric transmittance and radiance simulator. For the ORIS-drone path, assuming $d_2\ll d_1$ and that the entire $d_2$ path is subject to similar atmospheric conditions, we can apply the Beer-Lambert law for estimating the atmospheric loss as \cite{Ghassemlooy2012}
\begin{align}
\label{eq:loss2}
\tau _{\text{l,2}} =\exp\left(-\beta_{\text{l}}d_2\right),
\end{align}
where $\beta_{\text{l}}$ represents the atmospheric extinction coefficient. With the help of (\ref{eq:loss1}) and (\ref{eq:loss2}), the atmospheric loss over the HAP-ORIS-drone paths can be calculated as $\tau _{\text{l}}=\tau _{\text{l,1}}\tau _{\text{l,2}}$.

\begin{figure}[t]
\vspace{-0.5cm}
\centering
\includegraphics[scale=0.5]{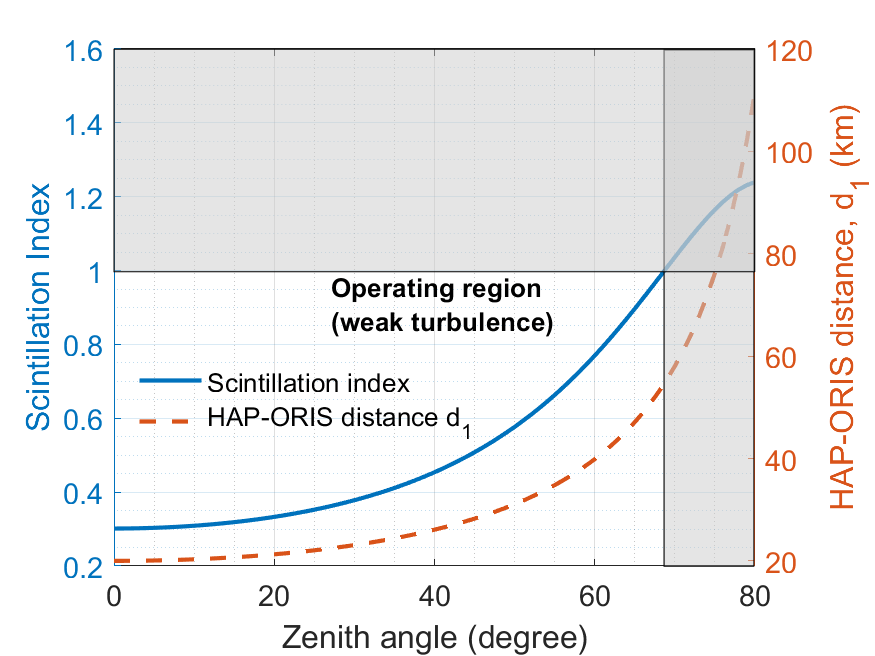}
\caption{Scintillation index versus HAP-ORIS distance $d_1$ and zenith angle $\varphi_{\text{i}}$. $\lambda\!\!=\!810$ nm, $A\!=\!\!3\!\times\!10^{-13}$ $\text{m}^{-2/3}$, $v_{\text{g}}\!=\!5$ m/s, $h_{\text{ORIS}}\!=\!50$ m, $h_{\text{HAP}}\!=\!20$ km, $R_{\text{E}}\!=\!6370$ km.}
\label{fig_2}
\vspace{-0.4cm}
\end{figure}

In describing $I_{\text{a}}$, we consider independent atmospheric turbulence-induced intensity fluctuations for the HAP-ORIS and ORIS-drone paths, denoted as $I_{\text{a},1}$ and $I_{\text{a},2}$, respectively. As a result, we have $I_{\text{a}}\!\!=\!\!I_{\text{a},1}I_{\text{a},2}$. In the weak turbulence regime, the log-normal PDT is adopted \cite{Andrews2005}, given by
\begin{align}
\label{eq:LN}
f\!\left ( I_{\text{a},\iota} \right )\!\!=\!\!\frac{1}{I_{\text{a},\iota}\sqrt{2\pi\sigma _{\text{R},\iota}^{2}}}\!\exp\!\!\left (\! \!-\frac{\left (\! \ln\!\left (  I_{\text{a},\iota} \right ) \!\!+\!\!\frac{\sigma _{\text{R},\iota}^{2}}{2}\!\right )^{\!2}}{2\sigma _{\text{R},\iota}^{2}}\! \right )\!\!,\!~\!\iota\!\!\in\!\!\{1,2\},\!
\end{align}
where $\sigma _{\text{R},\iota}^{2}$ denotes the Rytov variances for the HAP-ORIS (i.e., $\iota\!\!=\!\!1$) and ORIS-drone (i.e., $\iota\!\!=\!\!2$) paths over the atmosphere, generally expressed as \cite{Andrews2005}
\begin{align}
\label{eq:Rytov}
\sigma _{\text{R},\iota}^{2}\!\!=\!2.25k^{7/6}\!\sec^{11/6}\!\left ( \varphi _{\zeta } \right )\!\int_{\!h_{\text{ORIS}}}^{h_{\chi  }}\!C_{\text{n}}^{2}\!\left ( h \right )\!\!\left ( h\!-\! h_{\text{ORIS}}\right )^{5/6}\!\text{d}h,
\end{align}
where $\zeta\!\in\!\{\text{i},\text{r}\}$ and $\chi \!\in\!\{\text{HAP},\text{LAP}\}$ correspond to $\iota\!\in\!\{1,2\}$, respectively. Furthermore, $C_{\text{n}}^{2}\!\left ( h \right )$ denotes the refractive index structure parameter, which is determined from the Hufnagel-Valley model as \cite{Andrews2005}
\begin{align}
\label{Cn2_profile}
C_{\text{n}}^{2}\!\left ( h \right )\!=&0.00594\left ( \frac{v}{27} \right )^{\!2}\!\left ( 10^{-5}h \right )^{\!10}\exp\!\left (\! -\frac{h}{1000} \!\right ) \nonumber \\
&+2.7\!\times\!10^{-16}\exp\!\left (\! -\frac{h}{1500} \!\right )\!+\!A\exp\!\left (\! -\frac{h}{100} \!\right )\!,
\end{align}
where $h$ is the altitude in meters (m) and $A$ is the nominal value of $C_{\text{n}}^{2}\!\left ( 0 \right )$ at the ground in units of $\text{m}^{-2/3}$. Still referring to (\ref{Cn2_profile}), $v$ (m/s) is the root-mean-squared (rms) transverse wind speed at altitudes above 5 km, readily given by 
$v=\left ( \frac{1}{15000}\int_{5000}^{20000} \left [V\!\left ( h \right )  \right ]^{2} \text{d}h\right )^{\!1/2}$ \cite{Andrews2005},
where $V\!\left ( h \right )$ is the altitude-dependent Greenwood wind profile \cite{Greenwood1977}, appropriately modified to include $h_{\text{ORIS}}$ as 
$V\!\left ( h \right ) =v_{\text{g}}+30\exp\!\left [ -\left (\frac{h-12448+h_{\text{ORIS}}}{4800}   \right )^{2}\right ]$ \cite{Trinh2022},
where $v_{\text{g}}$ (m/s) denotes the ground wind speed. To quantify the turbulence strength, the scintillation index, defined as the normalized variance of $I_{\text{a}}$, is widely used. For a downlink path spanning from the HAP to ORIS, a general scintillation index, denoted as $\sigma _{I_{\text{a}}}^{2}$ applicable across all turbulence regimes, is given by \cite{Andrews2005}
\begin{align}
\sigma _{I_{\text{a}}}^{2}\!=\!\exp\!\!\left [\!\frac{0.49\sigma _{\text{R},1}^{2}}{\left (1\!+\!1.11\sigma _{\text{R},1}^{12/5}  \right )^{\!\!7/6}} \!+\! \frac{0.51\sigma _{\text{R},1}^{2}}{\left (1\!+\!0.69\sigma _{\text{R},1}^{12/5}  \right )^{\!\!5/6}}\!\right ]\!\!-\!\!1.
\end{align}
The scintillation index serves as a figure of merit indicating the strength of turbulence. Specifically, $\sigma _{I_{\text{a}}}^{2}<1$ indicates a weak turbulence regime, while $\sigma _{I_{\text{a}}}^{2}=1$ represents moderate turbulence, and $\sigma _{I_{\text{a}}}^{2}>1$ denotes strong turbulence conditions \cite{Ghalaii2022}. The scintillation index is investigated in Fig. \ref{fig_2} for the dominant HAP-ORIS path versus both the distance $d_1$ and the zenith angle $\varphi_{\text{i}}$. The HAP-ORIS distance $d_1$ varies with the zenith angle $\varphi_{\text{i}}$ and can be calculated as \cite{Ghalaii2022}
\begin{align}
d_1=&\sqrt{\left ( R_{\text{E}}+h_{\text{HAP}} \right )^{2}+\left ( R_{\text{E}}+h_{\text{ORIS}} \right )^{2}\!\left ( \cos^{2}\!\left ( \varphi _{\text{i}}\right ) -1\right )}\nonumber\\
&-\left ( R_{\text{E}}+h_{\text{ORIS}} \right )\!\cos\!\left ( \varphi _{\text{i}}\right ) ,
\end{align}
where $R_{\text{E}}$ denotes the Earth's radius. In particular, we consider daytime conditions along with $A\!=\!3\!\times\!10^{-13}$ in (\ref{Cn2_profile}) \cite{Andrews2005}, thereby ensuring that the scintillation index depicted in Fig. \ref{fig_2} represents the worst-case scenario for a quantum link. It transpires that $\sigma _{I_{\text{a}}}^{2}\!<\!1$ for $\varphi_{\text{i}}\!\leq\!68^{\circ}$, indicating that this range falls within the weak turbulence regime, which is favorable for quantum communications. However, for $\varphi_{\text{i}}\!>\!68^{\circ}$, the HAP-ORIS link enters the moderate-to-strong turbulence regime, significantly deteriorating the quantum signals and posing substantial challenges for link alignments due to the high zenith angles. Consequently, we restrict the QKD operations to the weak turbulence regime\footnote{We assume that $d_2\ll d_1$, which is logical given that the drone functions as a dynamic quantum platform capable of relocating to a favorable position to receive the reflected beam from ORIS. By maintaining $d_2$ within a reasonably short distance, such as less than 1 km, additional atmospheric attenuation losses are negligible. As a result, the scintillation index of the ORIS-drone path remains well below unity, staying within the weak turbulence regime.}, where $\varphi_{\text{i}}\!\leq\!68^{\circ}$ and $d_1\!\leq\! 53$ km, ensuring the validity of the log-normal PDT, as experimentally verified for quantum signals in \cite{Trinh2022}.

Finally, the GML coefficient $\tau_{\text{p}}$ includes both the deterministic geometrical loss, resulting from the truncation of the receiver aperture capturing only a portion of the optical beam's power, and the random PE-induced loss caused by drone hovering fluctuations. In ORIS-aided HAP-to-drone QKD links, accurately characterizing the GML is crucial, which is influenced by ORIS phase-shift profiles, atmospheric conditions, and PE due to drone hovering fluctuations. This challenge, unaddressed in the literature, is thoroughly investigated in Section \ref{sect:Section3} using the EHF principles.
\vspace{-0.5cm}
{\color{black}\subsection{Effects of ORIS on QKD Systems}
\subsubsection{ORIS Reflectance}
As mentioned in Section \ref{subsect:IA}, ORISs are classified into mirror-array-based and LC-based types. Mirror-array-based ORIS offers high contrast and fast response. However, it is limited by narrow beam deflection and low spatial resolution, making it less suitable for large-scale, cost-effective applications \cite{Ren2015}. In contrast, LC-based ORIS manipulates LC molecule orientation via independent electrodes, achieving high spatial resolution, wide beam deflection field of view, and polarization-independent reflectance \cite{Yu2024}. By leveraging established mass production techniques from the display industry, LC-based ORIS is ideal for scalable designs and continuous tunability. Assuming an ultra-thin LC metasurface with negligible inherent losses and a glass substrate cover introducing a total attenuation of $10\%$ for incident and reflected beams \cite{Alain2021}, the reflectance parameter for LC-based ORIS can be set to $\tau_{\text{ORIS}}\!=\!0.9$ in (\ref{eq:tau}).
\subsubsection{ORIS-Induced Delay Spread}
\begin{figure}[t]
\vspace{-0.1cm}
\centering
\includegraphics[scale=0.5]{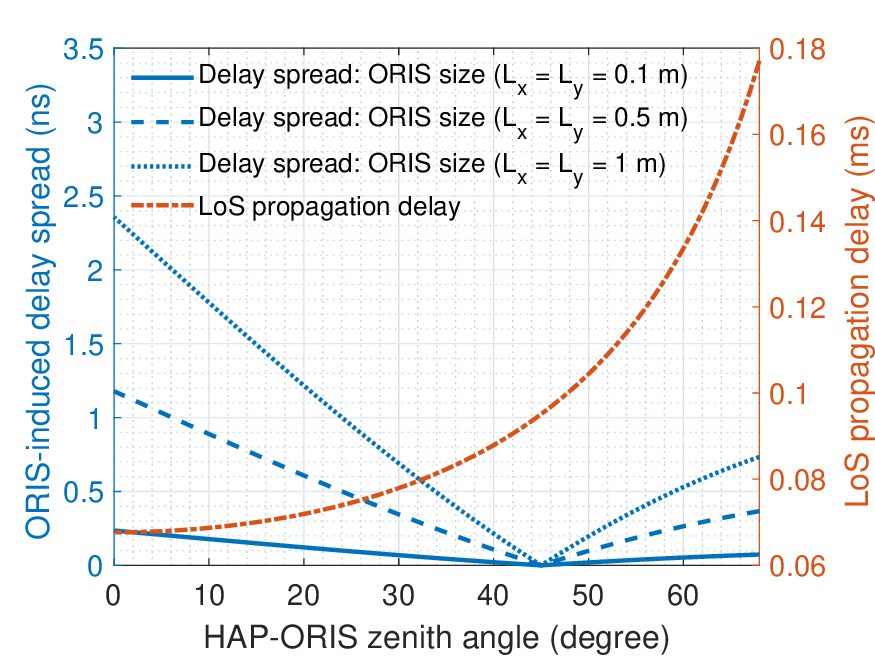}
\caption{{\color{black}ORIS-induced delay spread and LoS propagation delay versus HAP-ORIS zenith angle. $\theta_{\text{r}}\!=\!45^{\circ}$, $\lambda\!=\!810$ nm, $h_{\text{ORIS}}\!=\!50$ m, $h_{\text{HAP}}\!=\!20$ km, $h_{\text{LAP}}\!=\!300$ m, $R_{\text{E}}\!=\!6370$ km, $c=3\!\times\!10^{8}$ m/s.}}
\label{fig_3}
\vspace{-0.5cm}
\end{figure}
For a large ORIS, various elements reflect distinct portions of the incident beam, each traveling slightly different distances to the receiver. This variation can result in different delays across the reflected beam, leading to ORIS-induced delay spreads \cite{Ajam-arxiv}. The delay spread is defined as the difference between the maximum and minimum delay values across all ORIS elements. In high-data-rate classical FSO links (e.g., $1$–$10$ Gbps), the short symbol durations (e.g., $0.1$-$1$ ns) make the system susceptible to inter-symbol interference caused by such delay spreads. However, typical QKD systems, which operate at significantly lower rates with longer symbol durations (e.g., $5$–$10$ ns) \cite{Lu2022}, may be unaffected by ORIS-induced delay spreads. The delay profile of ORIS, denoted as $t_{\text{d}}(\mathbf{r}_\text{r})$, can be simplified to a linear model as 
%
$t_{\text{d}}(\mathbf{r}_\text{r})=t_{\text{LoS}}+t_{\text{ORIS}}$ \cite{Ajam-arxiv},
where 
\begin{align}
t_{\text{LoS}}=&\frac{d_1+d_2}{c}, \\
t_{\text{ORIS}}=&-\!\frac{\tan^{-1}\!\left (\! \frac{d_1}{z_{\text{R1}}} \!\right )}{kc}\!-\!\frac{x_{\text{r}}\!\left[\cos({\phi_{\text{r}}})\!\cos({\theta_{\text{r}}})\!+\!\cos({\phi_{\text{i}}})\!\cos({\theta_{\text{i}}})\right]}{c} \nonumber\\
&-\frac{y_{\text{r}}\!\left[\sin({\phi_{\text{i}}})\!\cos({\theta_{\text{i}}})\!+\!\sin({\phi_{\text{r}}})\!\cos({\theta_{\text{r}}})\right]}{c},
\end{align}
where $t_{\text{LoS}}$ denotes the end-to-end LoS propagation delay, $t_{\text{ORIS}}$ is the delay induced by ORIS, $c$ (m/s) is the light velocity, $z_{\text{R1}}\!=\!\frac{\pi w_0^2}{\lambda}$ is the Rayleigh range, which is determined by the beam waist at Tx $w_0\!=\!\lambda(\pi\theta_{\text{div}})^{-1}$ along with $\theta_{\text{div}}$ being the Tx half-angle beam divergence. Consequently, the delay spread can be written as
$\Delta t_{\text{d}}(\mathbf{r}_\text{r})=\max\left(t_{\text{d}}(\mathbf{r}_\text{r}) \right)\!-\!\min\left(t_{\text{d}}(\mathbf{r}_\text{r}) \right)$.
In Fig. \ref{fig_3}, $\Delta t_{\text{d}}(\mathbf{r}_\text{r})$ and $t_{\text{LoS}}$ are investigated with respect to zenith angle $\varphi_{\text{i}}$, considering the in-plane reflection, i.e., $\phi_{\text{i}}\!=\!0$ and $\phi_{\text{r}}\!=\!\pi$. It is confirmed that delay spreads increase with ORIS size and the difference between $\theta_{\text{i}}$ and $\theta_{\text{r}}$. For the ORIS with $L_x\!=\!L_y\!=\!1$ m (as in Section \ref{sect:Section3a}), the delay spreads remain below 2.4 ns, insufficient to cause pulse dispersion in typical QKD systems. When $\theta_{\text{i}}\!=\!\theta_{\text{r}}$, the delay spread is negligible, as all reflected components reach the receiver simultaneously.
\subsubsection{ORIS-Induced Polarization Changes}
\label{subsect:ORIS_pol}
\begin{figure}[t]
\vspace{-0.2cm}
\centering
\includegraphics[scale=0.5]{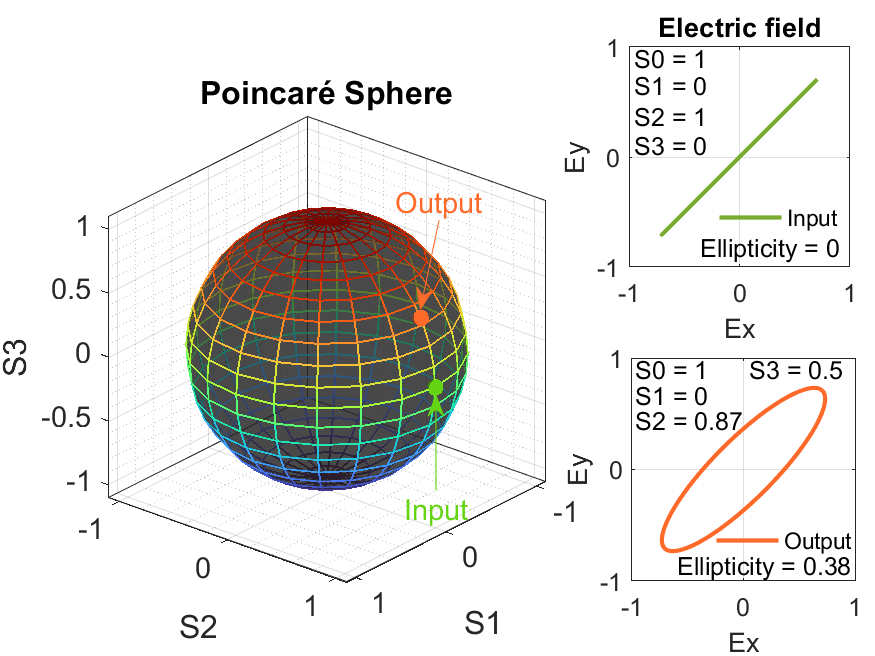}
\caption{{\color{black}ORIS-induced phase delay that converts linear polarization into elliptical polarization.}}
\label{fig_4}
\vspace{-0.4cm}
\end{figure}
The hovering nature of UAVs presents significant challenges for aligning the polarization in HAP-to-LAP QKD links \cite{Trinh2024,Takenaka2017}. These challenges are further exacerbated by ORIS, where varying reflection angles can misalign transmitted and received polarizations. For linear polarization, such variations induce phase delays between the orthogonal components, thereby transforming the linear polarization into elliptical polarization \cite{Alberto2016}. Fig. \ref{fig_4} illustrates this phenomenon, where a $45 ^{\circ}$-linearly polarized light reflected by the ORIS is transformed into elliptical polarization. This transformation is represented as a $30 ^{\circ}$ shift in the Stokes parameter S3 on the Poincaré sphere, resulting in the elliptical electric field. Such polarization alterations contribute to increased error rates in DV QKD systems due to inaccurate polarization detection\footnote{{\color{black}This is irrelevant in CV QKD, since information is encoded in the amplitude and phase quadratures of light, which are polarization-independent properties of the electromagnetic field.}}. To mitigate this, a motorized half-wave plate can be installed at the receiver to realign the polarization based on calculations from UAV inertial data and ORIS reflection angles \cite{Takenaka2017}. A quarter-wave plate can then restore elliptically polarized light to linear by eliminating the phase differences. The error probability associated with the efficiency of these corrections is quantified as the erroneous detector probability, $e_\text{det}$, included in Section \ref{subsect:DVQKD}.
\subsubsection{ORIS-Based Tracking Control}
The primary function of ORIS is to steer the optical beam toward the receiver and adaptively control the beam size to compensate for LAP hovering misalignments. ORIS is operated by a nearby ground base station via a high-speed optical fiber backhaul link. The HAP and LAP first align their gimbals toward the fixed coordinates of ORIS and transmit their inertial data to the base station. The base station calculates the incident and reflection angles based on the received data and generates the required phase-shift profile, which is sent to ORIS for implementation. {\color{black}The angle of arrival on the LAP can be determined by analyzing intensity differences detected by a quadrant detector with four regions} \cite{Alberto2022,Kolev2023,Alberto2024}, then transmitted to the base station for real-time control updates. Assuming the base station is close to ORIS and LAP, the propagation delays are negligible. To maintain alignment, ORIS must respond faster than the angle-of-arrival changes caused by LAP hovering, which typically have a frequency response below 50 Hz (i.e., equivalent to a 20 ms {\color{black}repetition} time) \cite{Trinh2021}. While recent advancements in ORIS technology achieve sub-ms response time \cite{Chang2023}, practical values may depend on various factors such as ORIS size and reflector design. If ORIS response time is longer than 20 ms, e.g., due to ORIS control or hardware impairments, the {\color{black}resultant} PEs can be quantified using the framework proposed in Section \ref{sect:Section3}, since any random displacements in horizontal and vertical axes may be modeled by i.n.i.d. Gaussian RVs \cite{Trinh2021}.
}
\vspace{-0.2cm}
\section{Statistical GML Based on EHF Principles}
\label{sect:Section3}
\subsection{Characterization of Optical Beams Using EHF Principles}
\label{sect:Section3a}
The impact of ORIS on the incident optical beam can be theoretically modeled using three different frameworks: (i) electromagnetic optics, which relies on vector field descriptions, (ii) wave optics, which uses scalar field descriptions, and (iii) geometric optics. The studies in \cite{Ajam2022,Ajam2024} found that wave optics provides accurate results for practical ORIS-aided FSO systems\footnote{{\color{black}The ORIS channel model based on wave optics applies to both classical and quantum systems, due to their shared use of coherent Gaussian laser beams, despite the lower power levels in quantum systems} \cite{Kundu2024}.}. Specifically, using scalar fields allows for explicit modeling of arbitrary ORIS phase-shift profiles, ORIS size, and Rx aperture size, which geometric optics fails to achieve. Consequently, the HF principles, based on a scalar-field analysis method, were applied to derive the beam reflected by the ORIS \cite{Ajam2022, Sipani2023, Ajam2024}. However, previous studies \cite{Ajam2022, Sipani2023, Ajam2024} overlooked that HF principles are only valid for optical beam propagation in a vacuum medium \cite{Lutomirski1971}. Particularly, the HF principles state that every point on a wavefront acts as a center of secondary disturbances, producing spherical wavelets, with the wavefront at a later instant becoming the envelope of these wavelets. For a random medium, the EHF principles state that the secondary wavefront is still determined by the envelope of spherical wavelets accruing from the primary wavefront. However, each wavelet is now influenced by the propagation of a spherical wave through the random turbulent medium \cite{Lutomirski1971}. Therefore, the key improvement of EHF principles lies in the characterization of turbulence-induced beam broadening and the random component of the complex phase of a spherical wave due to propagation in a turbulent medium \cite{Ricklin2002}.

Following \cite{Ajam2024} and applying the EHF principles \cite{Andrews2005, Lutomirski1971,Ricklin2002}, the electric field of the Gaussian laser beam incident on the ORIS can be expressed as
\begin{align}
\label{eq:E_i}
E_{\text{i}}\!\left ( \mathbf{r}_{\text{r}} \right )\!\!=\!\!C_{\text{i}}\!\exp\!\!\left(\!\!-\frac{x_{\text{r}}^{2}}{w_{\text{i},x}^{2}\!\!\left (d_1 \right )}\!\!-\!\!\frac{y_{\text{r}}^{2}}{w_{\text{i},y}^{2}\!\!\left ( d_1 \right )}\!\!-\!\!j\psi _{\text{i}}\!\left ( \mathbf{r}_{\text{r}} \right )\!\!+\!\! \Upsilon \!\left (\mathbf{r}_{\text{r}},d_1  \right )\!\!\right )\!\!,
\end{align}
with the phase $\psi _{\text{i}}\!\left ( \mathbf{r}_{\text{r}} \right )$ given by
\begin{align}
\psi _{\text{i}}\!\left ( \mathbf{r}_{\text{r}} \right )\!=\!k\left (\! \hat{d}_1\!+\!\frac{x_{\text{r}}^{2}\sin^{2}\!\left ( \theta _{\text{i}} \right )\!+\!y_{\text{r}}^{2}}{2R\!\left ( d_1 \right )} \!\right )\!-\!\tan^{-1}\!\left (\! \frac{d_1}{z_{\text{R1}}} \!\right ),
\end{align}
where we have $C_{\text{i}}\!=\!\sqrt{\frac{4\eta P_{\text{t}}\left | \sin\left ( \theta _{\text{i}} \right ) \right |}{\pi w^{2}(d_1) }}$ and $\eta$ is the channel impedance, $P_{\text{t}}$ is the transmitted power, and $w(d_1)$ is the beam waist at the distance $d_1$. Here, $\hat{d}_1\!=\!d_1\!-\!x_{\text{r}}\cos\!\left ( \theta _{\text{i}} \right )$. Furthermore, $w_{\text{i},x}\!=\!\frac{w(d_1)}{\sin\left(\theta _{\text{i}} \right)}$ and $w_{\text{i},x}\!=\!w(d_1)$ are the indcident beam widths on the ORIS plane in the $x$- and $y$-direction, respectively, while $\Upsilon \!\left (\mathbf{r}_{\text{r}},d_1  \right )$ denotes the phase pertubation of the field due to random inhomogeneities along the HAP-ORIS path. Subsequently, the electric field of the beam reflected by the ORIS and received at the Rx aperture of the drone can be written as
\begin{align}
\label{eq:E_r}
E_{\text{r}}\!\left ( \mathbf{r}' \right )\!=\!C_{\text{r}}\!&\iint_{\left ( x_{\text{r}},y_{\text{r}}\in\sum _{\text{ORIS}} \right )}\!E_{\text{i}}\!\left ( \mathbf{r}_{\text{r}} \right )\exp\!\left(-jk\left \| \mathbf{r}_{\text{o}}\!-\!  \mathbf{r}_{\text{r}} \right \|\right )\nonumber\\
&\times\!\exp\!\left [ -j\Phi _{\text{ORIS}}\!\left ( \mathbf{r}_{\text{r}}  \right ) \right ]\exp\!\left ( \Upsilon\!\left (\mathbf{r}',d_2  \right ) \right )\!{\color{black}\text{d}x_{\text{r}}\text{d}y_{\text{r}}},
\end{align}
where $C_{\text{r}}\!=\!\frac{\sqrt{\sin\left(\theta_{\text{r}}\right)}}{j\lambda d_2}$, $\mathbf{r}_{\text{o}}\!=\!\left (\mathbf{r}'\!+\!\mathbf{c}  \right )\mathbf{R}_{\text{rot}}$ with $\mathbf{r}'\!=\!\left( x',y',z'\right)$ being a point in the Rx aperture plane and $\mathbf{c}=\left ( 0,0,d_2 \right )$. 
Furthermore, we have a rotation matrix of
\begin{IEEEeqnarray}{rCL}
\mathbf{R}_{\text{rot}}=\begin{pmatrix}
-\sin\!\left ( \theta_{\text{r}} \right ) &0  &-\cos\!\left ( \theta_{\text{r}} \right ) \\
0&-1  &0 \\
-\cos\!\left ( \theta_{\text{r}} \right )& 0 &\sin\!\left ( \theta_{\text{r}} \right )
\end{pmatrix}, 
\end{IEEEeqnarray}
and $E_{\text{i}}\!\left ( \mathbf{r}_{\text{r}} \right )$ of (\ref{eq:E_r}) is given in (\ref{eq:E_i}), while $\Upsilon\!\left (\mathbf{r}',d_2  \right )$ denotes the phase perturbation of the field due to random inhomogeneities along the ORIS-drone path. 

In characterizing wave propagation through atmospheric turbulence, the statistical long-term-average moments of the optical field are of great interest. Thus, the mean electric field of $E_{\text{r}}\!\left ( \mathbf{r}' \right )$ can be written with the help of (\ref{eq:E_i}) and (\ref{eq:E_r}) as
\begin{align}
\label{eq:mean_E_r}
\left \langle E_{\text{r}}\!\left ( \mathbf{r}' \right ) \right \rangle\!\!=\!\!C_{\text{r}}C_{\text{i}}\!&\iint_{\!\left (\! x_{\text{r}},y_{\text{r}}\in\sum _{\text{ORIS}} \!\right )}\!\!\!\exp\!\!\left(\!\!-\frac{x_{\text{r}}^{2}}{w_{\text{i},x}^{2}\!\!\left (\!d_1\! \right )}\!-\!\frac{y_{\text{r}}^{2}}{w_{\text{i},y}^{2}\!\!\left (\! d_1\! \right )}\!-\!j\psi _{\text{i}}\!\left ( \mathbf{r}_{\text{r}}\right )\!\!\right )\nonumber\\
&\times\exp\!\left(-jk\left \| \mathbf{r}_{\text{o}}\!-\!  \mathbf{r}_{\text{r}} \right \|\right )\exp\!\left (\! -j\Phi _{\text{ORIS}}\!\left ( \mathbf{r}_{\text{r}}  \right ) \!\right )\!{\color{black}\text{d}x_{\text{r}}\text{d}y_{\text{r}}}\nonumber\\
&\times\left \langle \exp\!\left [ \Upsilon \!\left (\mathbf{r}_{\text{r}},d_1  \right )\right ]  \right \rangle\left \langle \exp\!\left [\Upsilon\!\left (\mathbf{r}',d_2  \right )\right ]  \right \rangle.
\end{align}
To calculate the ensemble averages appearing in (\ref{eq:mean_E_r}), we invoke the relationship \cite{Andrews2005}
\begin{align}
\left \langle \exp\!\left ( \Upsilon  \right ) \right \rangle\!=\!\exp\!\!\left [\! \left \langle \Upsilon  \right \rangle \!+\!\frac{1}{2}\!\left (\! \left \langle \Upsilon^{2} \right \rangle\!-\!\left \langle \Upsilon  \right \rangle^{\!2}\!\right )\!\right ]\!\!=\!\exp\!\left[E_1\!\!\left(0,0\right)\right],
\end{align}
which leads to the relationship with the second-order statistical moment of the optical field $E_1\!\left(0,0\right)$ that is real and independent of the observation point and the matrix elements between the input and output planes \cite{Andrews2005}. As a result, $\left \langle \exp\!\left [ \Upsilon \!\left (\mathbf{r}_{\text{r}},d_1  \right )\right ]  \right \rangle$ and $\left \langle \exp\!\left [\Upsilon\!\left (\mathbf{r}',d_2  \right )\right ]  \right \rangle$ in (\ref{eq:mean_E_r}) can be respectively formulated as \cite{Andrews2005}
\begin{align}
\label{eq:Upsilon1}
\left \langle \exp\!\left [ \Upsilon \!\left (\!\mathbf{r}_{\text{r}},d_1  \!\right )\right ]  \right \rangle\!\!&=\!\!\exp\!\!\left [\! -2\pi^2k^2\!\sec\!\left (\! \varphi_{\text{i}} \!\right )\!\!\int_{h_{\text{ORIS}}}^{h_{\text{HAP}}}\!\!\!\int_{0}^{\infty }\!\!\kappa \Phi _{n}\!\!\left (\! \kappa ,h \!\right )\!\right ]\!\!\text{d}\kappa \text{d}h, \\
\label{eq:Upsilon2}
\left \langle \exp\!\left [\Upsilon\!\left (\!\mathbf{r}',d_2  \!\right )\right ]  \right \rangle\!\!&=\!\!\exp\!\!\left [\! -2\pi^2k^2\!\sec\!\left (\! \varphi_{\text{r}} \!\right )\!\!\int_{h_{\text{ORIS}}}^{h_{\text{LAP}}}\!\!\!\int_{0}^{\infty }\!\!\kappa \Phi _{n}\!\!\left (\! \kappa ,h \!\right )\!\right ]\!\!\text{d}\kappa \text{d}h,
\end{align}
where $\Phi _{n}\!\left (\! \kappa ,h \!\right )$ denotes the spectral density\footnote{The general development is independent of the choice of spectrum model, which can be selected from the Kolmogorov, Tatarskii, von Karman, or exponential spectrums \cite{Andrews2005}.} of the refractive index fluctuations, with $\kappa$ being the scalar magnitude of the three-dimensional spatial wave number vector $\mathbf{K}$, under the assumption that the random medium is statistically homogeneous and isotropic in each transversal plane \cite{Andrews2005}. It is observed that the mean fields $\left \langle \exp\!\left [ \Upsilon \!\left (\!\mathbf{r}_{\text{r}},d_1 \! \right )\right ]  \right \rangle$ and $\left \langle \exp\!\left [\Upsilon\!\left(\!\mathbf{r}',d_2  \!\right )\right ]  \right \rangle$ in (\ref{eq:Upsilon1}) and (\ref{eq:Upsilon2}), respectively, tend to zero for both visible and infrared wavelengths due to the $k^2$ term. Hence, phase perturbation becomes negligible for our QKD links at $\lambda\!=\!810$ nm.

To this end, the effect of atmospheric turbulence reduces to beam broadening in both the HAP-ORIS and ORIS-drone links, which was not considered in previous studies \cite{Ajam2022, Sipani2023, Ajam2024}. The beam broadening effect of the HAP-ORIS path can be characterized by $w(d_1)$, which represents the long-term beam waist at a distance $d_1$ after propagating through the atmosphere. {\color{black}For} a collimated beam, $w(d_1)$ is given by \cite{Andrews2005}
\begin{align}
\label{eq:w_d1}
w(d_1)\!=\!w_0\sqrt{\left (1\!+\!\Lambda_0^{2}  \right )\left ( 1\!+\!T \right )},
\end{align}
where we have $\Lambda_0\!=\!\frac{2d_1}{kw_0^{2}}$, and $T$ characterizes the turbulence-induced beam broadening effect, expressed as \cite{Andrews2005}
\begin{align}
\label{eq:T}
T=&4.35\Lambda ^{5/6} k^{7/6}\left (h_{\text{HAP}}-h_{\text{ORIS}} \right )^{5/6}\sec^{11/6}\!\left ( \varphi_{\text{i}} \right )\nonumber\\
&\times\int_{h_{\text{ORIS}}}^{h_{\text{HAP}}}C_{\text{n}}^{2}\!\left ( h \right )\!\left ( \frac{h-h_{\text{ORIS}}}{h_{\text{HAP}}-h_{\text{ORIS}}} \right )^{\!5/3}\text{d}h,
\end{align}
where $\Lambda=\frac{\Lambda_0}{1+\Lambda_0^{2}}$. Upon using the parameters provided in the caption of Fig. \ref{fig_2} along with $\theta_{\text{div}}=16.5$ $\mu$rad for a collimated beam \cite{Trinh2022}, and substituting them into (\ref{eq:w_d1}) and (\ref{eq:T}), we find that the maximum beam widths incident on the ORIS are $w_{\text{i},x}\!=\!36.81$ cm and $w_{\text{i},y}\!=\!34.14$ cm at $\varphi_{\text{i}}\!=\!68^{\circ}$. This results in an elliptical beam with major and minor diameters of 73.62 m and 68.28 m, respectively. Given that the ORIS is fixed on the building rooftop, the Tx is equipped with an accurate pointing system \cite{Alberto2024}, and the ORIS size exceeds the beam footprint, the PE in the HAP-ORIS link thus can be considered negligible. To ensure that the incident beam footprint remains well within the large ORIS, the dimensions of $\sum_{\text{ORIS}}$ in this paper should be $L_x\!=\!L_y\!=\!1$ m. For the ORIS-drone path, the turbulence-induced beam broadening effect must be considered when calculating the beam widths at the Rx aperture for different ORIS phase-shift profiles. This is discussed in Section \ref{sect:Section3b}.
\vspace{-0.4cm}
\subsection{GML With ORIS Phase-Shift Profiles and Drone Hovering Fluctuations}
\label{sect:Section3b}
The GML coefficient $\tau_{\text{p}}$ is defined as \cite{Ajam2024}
\begin{align}
\label{eq:tau_p}
\tau_{\text{p}}=\frac{1}{2\eta P_{\text{t}}}\iint_{\mathcal{A}_{\text{Rx}}}\left | \left \langle E_{\text{r}}\!\left ( \mathbf{r}' \right ) \right \rangle \right |^{2}\text{d}\mathcal{A}_{\text{Rx}},
\end{align}
where $\mathcal{A}_{\text{Rx}}$ denotes the area of the Rx aperture and $\left \langle E_{\text{r}}\!\left ( \mathbf{r}' \right ) \right \rangle$ is given in (\ref{eq:mean_E_r}). Following the approach in \cite{Ajam2024}, we derive closed-form solutions for (\ref{eq:tau_p}) to estimate $\tau_{\text{p}}$, which depends on different ORIS phase-shift profiles and $\mathcal{A}_{\text{Rx}}$. Assuming that $\sum_{\text{ORIS}}\gg A_{\text{in}}$, where $A_{\text{in}}=\pi w_{\text{i},x}w_{\text{i},y}$ represents the area of the equivalent beam footprint incident on the ORIS, $\tau_{\text{p}}$ follows the saturated power scaling regime \cite{Ajam2024}, where all the power of the incident beam on the ORIS is reflected towards the drone. Thus, $\tau_{\text{p}}$ is independent of $\sum_{\text{ORIS}}$ and characterized by the GML governed by the Rx beam footprint, $\mathcal{A}_{\text{Rx}}$, and the average PE loss imposed by drone hovering fluctuations.

\begin{lemma}
\label{Lemma1}
Using the LPS profile in (\ref{eq:LP}) and assuming that the hovering fluctuations in positions of the drone in $x'$ and $y'$ axes, respectively denoted as $\tilde{x}'$ and $\tilde{y}'$, are i.n.i.d. Gaussian RVs, i.e., $\tilde{x}'\!\sim\! \mathcal{N}(\mu_{\tilde{x}'},\sigma_{\tilde{x}'}^2)$ and $\tilde{y}'\!\sim\! \mathcal{N}(\mu_{\tilde{y}'},\sigma_{\tilde{y}'}^2)$, the statistical average GML coefficient $\tau_{\text{p}}$ can be approximated by $\tau_{\text{p}}^{\text{LPS}}$, as given in (\ref{eq:LP_CL}), {\color{black}where} $a$ denotes the radius of the Rx aperture, while $w_{\text{rx},x'}^{\text{LPS}}\!=\!w\!\left ( d_1\right )\!\frac{\left | \sin\left ( \theta_{\text{r}}  \right )\right|}{\left | \sin\left ( \theta_{\text{i}}  \right )\right| } \sqrt{\varepsilon\left ( \frac{\sin^{2}\left ( \theta_{\text{i}}  \right )}{\sin^{2}\left ( \theta_{\text{r}}  \right )}\Lambda _{1}  \right )^{\!2}\!+\!1}$ and $w_{\text{rx},y'}^{\text{LPS}}\!=\!w\!\left ( d_1\right ) \!\sqrt{\varepsilon\Lambda _{1}^{2}\!+\!1}$ are the equivalent beam widths given by the LPS profile at the Rx aperture in the $x'$ and $y'$ axes, respectively. Finally, $\Lambda _{1} \!=\!\frac{2d_2}{kw^2(d_1)}$ and $\varepsilon=1\!+\!\frac{2w^2(d_1)}{\rho_0^2}$, with $\rho_0\!=\!\left (1.45k^2\!\int_{h_{\text{ORIS}}}^{h_{\text{LAP}}} C_{\text{n}}^{2}\!\left ( h \right ) \text{d}h\right )^{\!-3/5}\!\cos^{3/5}\!\left (\varphi _{\text{r}}  \right )$.
\begin{figure*}[b]
\vspace{-0.4cm}
\noindent\makebox[\linewidth]{\rule{\textwidth}{0.4pt}}
\begin{align}
\label{eq:LP_CL}
 {\color{black}\left \langle \tau_{\text{p}}^{\text{LPS}}\right \rangle}\!=\!\frac{1}{4}\!\left [\! \text{erf}\!\left (\! \frac{\frac{{\color{black}\sqrt{2}}\mu_{\tilde{x}'}}{w_{\text{rx},x'}^{\text{LPS}}}\!+\!\frac{a\sqrt{\pi}}{{\color{black}\sqrt{2}}w_{\text{rx},x'}^{\text{LPS}}}}{\sqrt{1\!+\!\frac{{\color{black}4}\sigma_{\tilde{x}'}^2}{\left(\!w_{\text{rx},x'}^{\text{LPS}}\!\right)^{\!2}}}} \!\right )\!\!\!-\!\text{erf}\!\left (\! \frac{\frac{{\color{black}\sqrt{2}}\mu_{\tilde{x}'}}{w_{\text{rx},x'}^{\text{LPS}}}\!-\!\frac{a\sqrt{\pi}}{{\color{black}\sqrt{2}}w_{\text{rx},x'}^{\text{LPS}}}}{\sqrt{1\!+\!\frac{{\color{black}4}\sigma_{\tilde{x}'}^2}{\left(\!w_{\text{rx},x'}^{\text{LPS}}\!\right)^{\!2}}}}\!\right ) \!\right ]\!\!\left [\! \text{erf}\!\left (\! \frac{\frac{{\color{black}\sqrt{2}}\mu_{\tilde{y}'}}{w_{\text{rx},y'}^{\text{LPS}}}\!+\!\frac{a\sqrt{\pi}}{{\color{black}\sqrt{2}}w_{\text{rx},y'}^{\text{LPS}}}}{\sqrt{1\!+\!\frac{{\color{black}4}\sigma_{\tilde{y}'}^2}{\left(\!w_{\text{rx},y'}^{\text{LPS}}\!\right)^{\!2}}}} \!\right )\!\!\!-\!\text{erf}\!\left (\! \frac{\frac{{\color{black}\sqrt{2}}\mu_{\tilde{y}'}}{w_{\text{rx},y'}^{\text{LPS}}}\!-\!\frac{a\sqrt{\pi}}{{\color{black}\sqrt{2}}w_{\text{rx},y'}^{\text{LPS}}}}{\sqrt{1\!+\!\frac{{\color{black}4}\sigma_{\tilde{y}'}^2}{\left(\!w_{\text{rx},y'}^{\text{LPS}}\!\right)^{\!2}}}}\!\right ) \!\right ]\!\!.
\end{align}
\end{figure*}
\end{lemma}
\begin{proof}
See Appendix \ref{appendix_A}.
\end{proof}
\begin{remark}
\label{Remark1}
From Lemma \ref{Lemma1}, we observe that $\Lambda _{1}$ quantifies the increase in beam width along the ORIS-drone path at a distance $d_2$ solely caused by diffraction. In addition, $\varepsilon$ characterizes the beam broadening due to atmospheric turbulence after reflection by the ORIS. In the absence of atmospheric turbulence, i.e., in a vacuum, $\rho_0\rightarrow\infty$ and $\varepsilon\rightarrow1$. Hence, under these conditions, the beam broadening is governed purely by free-space diffraction, as described by $\Lambda _{1}$. {\color{black}Additionally, in the absence of drone hovering fluctuations, i.e., $\mu_{\tilde{x}'}\rightarrow 0$, $\sigma_{\tilde{y}'}^2 \rightarrow 0$, (\ref{eq:LP_CL}) reduces to \cite[(22)]{Ajam2024}.}
\end{remark}
\begin{lemma}
\label{Lemma2}
Using the QPS profile in (\ref{eq:QP}) and assuming that the hovering fluctuations in positions of the drone in $x'$ and $y'$ axes, respectively denoted as $\tilde{x}'$ and $\tilde{y}'$, are i.n.i.d. Gaussian RVs, i.e., $\tilde{x}'\!\sim\! \mathcal{N}(\mu_{\tilde{x}'},\sigma_{\tilde{x}'}^2)$ and $\tilde{y}'\!\sim\! \mathcal{N}(\mu_{\tilde{y}'},\sigma_{\tilde{y}'}^2)$, the statistical average GML coefficient $\tau_{\text{p}}$ can be approximated by $\tau_{\text{p}}^{\text{QPS}}$, as given in (\ref{eq:QP_CL}), {\color{black}where} $w_{\text{rx},x'}^{\text{QPS}}\!=\!w\!\left ( d_1\right )\!\frac{\left|\sin\left ( \theta_{\text{r}}  \right )\right|}{\left|\sin\left ( \theta_{\text{i}}  \right )\right|} \sqrt{\varepsilon\!\left ( \frac{\sin^{2}\left ( \theta_{\text{i}}  \right )}{\sin^{2}\left ( \theta_{\text{r}}  \right )}\Lambda _{1}  \right )^{\!2}\!+\!\left ( \frac{d_2}{2f} \right )^{\!2}}$ and $w_{\text{rx},y'}^{\text{QPS}}\!=\!w\!\left ( d_1\right ) \!\sqrt{\varepsilon\Lambda _{1}^{2}\!+\!\left ( \frac{d_2}{2f} \right )^{\!2}}$ are the equivalent beam widths induced by the QPS profile at the Rx aperture in the $x'$ and $y'$ axes, respectively. Finally, $a$, $\varepsilon$, and $\rho_0$ are defined in Lemma \ref{Lemma1}.
\begin{figure*}[b]
\vspace{-0.4cm}
\noindent\makebox[\linewidth]{\rule{\textwidth}{0.4pt}}
\begin{align}
\label{eq:QP_CL}
 {\color{black}\left \langle\tau_{\text{p}}^{\text{QPS}}\right \rangle}\!=\!\frac{1}{4}\!\left [\! \text{erf}\!\left (\! \frac{\frac{{\color{black}\sqrt{2}}\mu_{\tilde{x}'}}{w_{\text{rx},x'}^{\text{QPS}}}\!+\!\frac{a\sqrt{\pi}}{{\color{black}\sqrt{2}}w_{\text{rx},x'}^{\text{QPS}}}}{\sqrt{1\!+\!\frac{{\color{black}4}\sigma_{\tilde{x}'}^2}{\left(\!w_{\text{rx},x'}^{\text{QPS}}\!\right)^{\!2}}}} \!\right )\!\!\!-\!\text{erf}\!\left (\! \frac{\frac{{\color{black}\sqrt{2}}\mu_{\tilde{x}'}}{w_{\text{rx},x'}^{\text{QPS}}}\!-\!\frac{a\sqrt{\pi}}{{\color{black}\sqrt{2}}w_{\text{rx},x'}^{\text{QPS}}}}{\sqrt{1\!+\!\frac{{\color{black}4}\sigma_{\tilde{x}'}^2}{\left(\!w_{\text{rx},x'}^{\text{QPS}}\!\right)^{\!2}}}}\!\right ) \!\right ]\!\!\left [\! \text{erf}\!\left (\! \frac{\frac{{\color{black}\sqrt{2}}\mu_{\tilde{y}'}}{w_{\text{rx},y'}^{\text{QPS}}}\!+\!\frac{a\sqrt{\pi}}{{\color{black}\sqrt{2}}w_{\text{rx},y'}^{\text{QPS}}}}{\sqrt{1\!+\!\frac{{\color{black}4}\sigma_{\tilde{y}'}^2}{\left(\!w_{\text{rx},y'}^{\text{QPS}}\!\right)^{\!2}}}} \!\right )\!\!\!-\!\text{erf}\!\left (\! \frac{\frac{{\color{black}\sqrt{2}}\mu_{\tilde{y}'}}{w_{\text{rx},y'}^{\text{QPS}}}\!-\!\frac{a\sqrt{\pi}}{{\color{black}\sqrt{2}}w_{\text{rx},y'}^{\text{QPS}}}}{\sqrt{1\!+\!\frac{{\color{black}4}\sigma_{\tilde{y}'}^2}{\left(\!w_{\text{rx},y'}^{\text{QPS}}\!\right)^{\!2}}}}\!\right ) \!\right ]\!\!.
\end{align}
\vspace{-0.5cm}
\end{figure*}
\end{lemma}
\begin{proof}
See Appendix \ref{appendix_B}.
\end{proof}
\begin{remark}
\label{Remark2}
Lemma \ref{Lemma2} reveals that increasing the focus distance $f$ results in a smaller beam footprint at the Rx aperture plane. Consequently, by adaptively adjusting $f$, the beam width at the receiver can be optimized for enhancing the performance under varying PE severities caused by drone hovering fluctuations. In the absence of atmospheric turbulence-induced beam broadening {\color{black}and drone hovering fluctuations,} (\ref{eq:QP_CL}) reduces to \cite[(24)]{Ajam2024}. Additionally, by comparing (\ref{eq:QP_CL}) and (\ref{eq:LP_CL}), we find that setting $f\!=\!d_2/2$ causes an ORIS with a QPS profile to behave identically to one with a LPS profile. 
\end{remark}
\begin{lemma}
\label{Lemma3}
Using the FPS profile in (\ref{eq:FP}) and assuming that the hovering fluctuations in positions of the drone in $x'$ and $y'$ axes, respectively denoted as $\tilde{x}'$ and $\tilde{y}'$, are i.n.i.d. Gaussian RVs, i.e., $\tilde{x}'\!\sim\! \mathcal{N}(\mu_{\tilde{x}'},\sigma_{\tilde{x}'}^2)$ and $\tilde{y}'\!\sim\! \mathcal{N}(\mu_{\tilde{y}'},\sigma_{\tilde{y}'}^2)$, the statistical average GML coefficient $\tau_{\text{p}}$ can be approximated by $\tau_{\text{p}}^{\text{FPS}}$, as given in (\ref{eq:FP_CL}), {\color{black}where} $w_{\text{rx},x'}^{\text{FPS}}\!\!=\!\!w\!\left ( d_1\right )\!\!\frac{\left|\sin\left ( \theta_{\text{i}}  \right )\right|}{\left|\sin\left ( \theta_{\text{r}}  \right )\right|}\sqrt{\varepsilon\Lambda_1^2}$ and $w_{\text{rx},y'}^{\text{FPS}}\!\!=\!\!w\!\left ( d_1\right )\!\!\sqrt{\varepsilon\Lambda_1^2}$ are the equivalent beam widths induced by the FPS profile at the Rx aperture in the $x'$ and $y'$ axes, respectively. {\color{black}Moreover}, $a$, $\varepsilon$, and $\rho_0$ are defined in Lemma \ref{Lemma1}.
\begin{figure*}[b]
\vspace{-0.2cm}
\begin{align}
\label{eq:FP_CL}
{\color{black}\left \langle\tau_{\text{p}}^{\text{FPS}}\right \rangle}\!=\!\frac{1}{4}\!\left [\! \text{erf}\!\left (\! \frac{\frac{{\color{black}\sqrt{2}}\mu_{\tilde{x}'}}{w_{\text{rx},x'}^{\text{FPS}}}\!+\!\frac{a\sqrt{\pi}}{{\color{black}\sqrt{2}}w_{\text{rx},x'}^{\text{FPS}}}}{\sqrt{1\!+\!\frac{{\color{black}4}\sigma_{\tilde{x}'}^2}{\left(\!w_{\text{rx},x'}^{\text{FPS}}\!\right)^{\!2}}}} \!\right )\!\!\!-\!\text{erf}\!\left (\! \frac{\frac{{\color{black}\sqrt{2}}\mu_{\tilde{x}'}}{w_{\text{rx},x'}^{\text{FPS}}}\!-\!\frac{a\sqrt{\pi}}{{\color{black}\sqrt{2}}w_{\text{rx},x'}^{\text{FPS}}}}{\sqrt{1\!+\!\frac{{\color{black}4}\sigma_{\tilde{x}'}^2}{\left(\!w_{\text{rx},x'}^{\text{FPS}}\!\right)^{\!2}}}}\!\right ) \!\right ]\!\!\left [\! \text{erf}\!\left (\! \frac{\frac{{\color{black}\sqrt{2}}\mu_{\tilde{y}'}}{w_{\text{rx},y'}^{\text{FPS}}}\!+\!\frac{a\sqrt{\pi}}{{\color{black}\sqrt{2}}w_{\text{rx},y'}^{\text{FPS}}}}{\sqrt{1\!+\!\frac{{\color{black}4}\sigma_{\tilde{y}'}^2}{\left(\!w_{\text{rx},y'}^{\text{FPS}}\!\right)^{\!2}}}} \!\right )\!\!\!-\!\text{erf}\!\left (\! \frac{\frac{{\color{black}\sqrt{2}}\mu_{\tilde{y}'}}{w_{\text{rx},y'}^{\text{FPS}}}\!-\!\frac{a\sqrt{\pi}}{{\color{black}\sqrt{2}}w_{\text{rx},y'}^{\text{FPS}}}}{\sqrt{1\!+\!\frac{{\color{black}4}\sigma_{\tilde{y}'}^2}{\left(\!w_{\text{rx},y'}^{\text{FPS}}\!\right)^{\!2}}}}\!\right ) \!\right ]\!\!.
\end{align}
\vspace{-0.5cm}
\end{figure*}
\end{lemma}
\begin{proof}
See Appendix \ref{appendix_C}.
\end{proof}
\begin{remark}
\vspace{-0.1cm}
\label{Remark3}
Lemma \ref{Lemma3} demonstrates that the beam footprint at the receiver is significantly smaller than those described in Lemmas \ref{Lemma1} and \ref{Lemma2}. In the absence of atmospheric turbulence-induced beam broadening {\color{black}and drone hovering fluctuations,} (\ref{eq:FP_CL}) simplifies to \cite[(25)]{Ajam2024} and the beam footprint is on the order of $w_0$, which is much smaller than the Rx aperture radius $a$, leading to $\tau_{\text{p}}^{\text{FPS}}\approx0$ dB. Furthermore, by comparing (\ref{eq:FP_CL}) and (\ref{eq:QP_CL}), we find that setting $f\!=\!\infty$ in (\ref{eq:QP_CL}) makes an ORIS with a QPS profile behave identically to one with a FPS profile.
\end{remark}
\vspace{-0.4cm}
\section{{\color{black}Applications in QKD Systems}}
\label{sect:Section4}
\subsection{Bounding the SKR of QKD Systems}
Optical communications over free-space links inherently experience channel impairments, which can be typically characterized by the transmissivity coefficient $\tau$ defined in (\ref{eq:tau}). For a lossy channel having arbitrary transmissivity $\tau$, the PLOB bound establishes the ultimate information-theoretic upper limit for the SKR of any DV/CV-QKD protocols \cite{Kundu2024,Ghalaii2022,Pirandola2017}, given by
\begin{align}
\label{eq:PLOB}
\mathfrak{R}\!\leq\!-\log_{2}\!\left(1-\tau\right)\!=\!-\log_{2}\!\left(1-\tau _{\text{eff}}{\color{black}\tau _{\text{ORIS}}}\tau _{\text{l}}I_{\text{a}}\tau_{\text{p}}\right).
\end{align}
{\color{black}
\begin{corollary}
\vspace{-0.5cm}
\label{Corollary1}
The instantaneous $\tau_{\text{p}}$ for the LPS and QPS can be approximated as
\begin{align}
\label{eq:tau_p_U}
\tau_{\text{p}}^{\text{U}}\!\approx\!A_{x'}A_{y'}\!\exp\!\left(\!-\frac{2\tilde{x}'^{2}}{(w_{\text{rx},x'\!(\text{eq})}^{\text{U}})^{2}}\!\right)\!\exp\!\left(\!-\frac{2\tilde{y}'^{2}}{(w_{\text{rx},y'\!(\text{eq})}^{\text{U}})^{2}}\!\right),
\end{align}
where $A_{x'}\!=\!\erf\!\!\left(\!\!\frac{a\sqrt{\pi}}{{\color{black}\sqrt{2}}w_{\text{rx},x'}^{\text{U}}}\!\!\right)$,$A_{y'}\!=\!\erf\!\!\left(\!\!\frac{a\sqrt{\pi}}{{\color{black}\sqrt{2}}w_{\text{rx},y'}^{\text{U}}}\!\!\right)$, $\text{U}\!\!\in\!\!\{\text{LPS},\text{QPS}\}$. Hence, the statistical average of $\tau_{\text{p}}^{\text{U}}$ can be written as
\begin{align}
\label{eq:GML_approx}
\left \langle\tau_{\text{p}}^{\text{U}}\right \rangle\!\!=&\frac{A_{x'}A_{y'}\gamma_{x'}\gamma_{y'}}{\sqrt{(1\!+\!\gamma_{x'}^{2})(1\!+\!\gamma_{y'}^{2})}} \nonumber\\
&\!\times\!\exp\!\!\left(\!-\frac{2}{w_{\text{rx},x'\!(\text{eq})}^{\text{U}}w_{\text{rx},y'\!(\text{eq})}^{\text{U}}}\!\!\left[\frac{\mu_{\tilde{x}'}^{2}}{1\!+\!\frac{1}{\gamma_{x'}^{2}}}\!+\!\frac{\mu_{\tilde{y}'}^{2}}{1\!+\!\frac{1}{\gamma_{y'}^{2}}}\right]\!\right)\!\!,
\end{align}
where $w_{\text{rx},x'\!(\text{eq})}^{\text{U}}\!=\!w_{\text{rx},x'}^{\text{U}}\sqrt{\frac{\sqrt{\pi}A_{x'}}{2\nu_{x'}\exp\!(-\nu_{x'}^{2})}}$ and $w_{\text{rx},y'\!(\text{eq})}^{\text{U}}\!=\!w_{\text{rx},y'}^{\text{U}}\sqrt{\frac{\sqrt{\pi}A_{y'}}{2\nu_{y'}\exp\!(-\nu_{y'}^{2})}}$ are the equivalent beam widths in $x'$ and $y'$ axes, respectively, with $\nu_{x'}\!=\!\frac{a\sqrt{\pi}}{{\color{black}\sqrt{2}}w_{\text{rx},x'}^{\text{U}}}$ and $\nu_{y'}\!=\!\frac{a\sqrt{\pi}}{{\color{black}\sqrt{2}}w_{\text{rx},y'}^{\text{U}}}$. Moreover, $\gamma_{x'}\!=\!\frac{w_{\text{rx},x'\!(\text{eq})}^{\text{U}}}{2\sigma_{\tilde{x}'}}$ and $\gamma_{y'}\!=\!\frac{w_{\text{rx},y'\!(\text{eq})}^{\text{U}}}{2\sigma_{\tilde{y}'}}$.
\end{corollary}
\begin{proof}
See Appendix \ref{appendix_D}.
\end{proof}
\begin{remark}
\label{Remark4}
From Corollary \ref{Corollary1}, the exact expressions of $\left \langle\tau_{\text{p}}^{\text{LPS}}\right \rangle$ and $\left \langle\tau_{\text{p}}^{\text{QPS}}\right \rangle$ in (\ref{eq:LP_CL}) and (\ref{eq:QP_CL}) can be approximated by (\ref{eq:GML_approx}). The approximation is valid when $w_{\text{rx},x'}^{\text{U}}$ and $w_{\text{rx},y'}^{\text{U}}$ are larger than the Rx aperture radius $a$, with $w_{\text{rx},x'}^{\text{U}},w_{\text{rx},y'}^{\text{U}}\geq6a$ achieving negligible approximation errors \cite{Farid2007}. Consequently, Corollary \ref{Corollary1} only applies to LPS and QPS profiles. It is noted that the derivation of (\ref{eq:tau_p_U}) is useful for the formulation of Corollary \ref{Corollary2}.
\end{remark}
}
The average PLOB bound of the SKR $\mathfrak{R}$ over the HAP-ORIS and ORIS-drone links can be expressed as
\begin{align}
\label{eq:PLOB_int}
{\color{black}\left \langle\mathfrak{R}\right \rangle\!\leq\!\int_{0}^{1}\!-\log_{2}\!\left(1\!-\!\tau \right)\!f\!\left (\tau \right )\!\text{d}\tau}
\end{align}
where {\color{black}$f\!\left (\tau \right )$ is the PDT of the transmissivity coefficient $\tau$.}
\begin{corollary}
\label{Corollary2}
A closed-form expression of the average PLOB bound of the SKR in (\ref{eq:PLOB_int}), {\color{black}considering that $I_{\text{a}}$ and $\tau_{\text{p}}$ are independent RVs in (\ref{eq:PLOB})}, is given by {\color{black}(\ref{eq:PLOB_CL}),}
\begin{figure*}[b]
\vspace{-0.4cm}
\noindent\makebox[\linewidth]{\rule{\textwidth}{0.4pt}}
\begin{align}
\label{eq:PLOB_CL}
{\color{black}\left \langle\mathfrak{R}\right \rangle\!\!\leq\!\!\frac{\gamma_{\text{mod}}^{2}\sigma _{\text{R}}}{\sqrt{2}}\!\exp\!\!\left (\!\! -\frac{\sigma _{\text{R}}^{2}}{2} \gamma_{\text{mod}}^{4}\!\right )\!\!\sum_{g =1 }^{G}\!-w _{g }\text{erfc}\!\left( x_{g } \right)\!\exp\!\!\left (\! x_{g }^{2}\!+\!\sqrt{2}\sigma _{\text{R}}\gamma_{\text{mod}}^{2}x_{g } \!\right )\!\!\log_{2}\!\!\Bigg[\!1\!\!-\!\!A_{\text{mod}}\tau _{\text{eff}}\tau _{\text{ORIS}}\tau _{\text{l}}\!\exp\!\!\left (\!\! \sqrt{2}\sigma _{\text{R}} x_{g }\!-\!\frac{\sigma _{\text{R}}^{2}}{2}\!\!\left (\! 1\!+\!2\gamma_{\text{mod}}^{2} \right ) \!\!\!\right )\!\!\!\Bigg]\!\!.}
\end{align}
\vspace{-0.5cm}
\end{figure*}
where $G$ is the Gauss-{\color{black}Hermite} polynomial order, $w _{g}$ and $x_{g}$ are the weight factors and the abscissas of the Gauss-{\color{black}Hermite} quadrature, respectively \cite[{\color{black}Table 25.10}]{Abra1972}. {\color{black}$A_\text{mod}\!\!=\!\!A_{x'}A_{y'}\Psi$ with $\Psi\!=\!\exp\!\Big(\! \frac{1}{\gamma_\text{mod}^{2}}\!-\!\frac{1}{2\gamma _{x'}^{2}}\!-\!\frac{1}{2\gamma_{y'}^{2}}\!-\!\frac{\mu _{\tilde{x}'}^2}{2\sigma _{\tilde{x}'}^{2}\gamma _{x'}^{2}} \!-\!\frac{\mu _{\tilde{y}'}^2}{2\sigma _{\tilde{y}'}^{2}\gamma _{y'}^{2}}  \! \Big )$. $\gamma_\text{mod}\!=\!\frac{\sqrt{w_{\text{rx},x'\!(\text{eq})}^{\text{U}}w_{\text{rx},y'\!(\text{eq})}^{\text{U}}}}{2\sigma _\text{mod}}$ with $\sigma _\text{mod}\!=\!\Big ( \frac{3\mu_{\tilde{x}'}^{2}\sigma _{\tilde{x}'}^{4}+3\mu_{\tilde{y}'}^{2}\sigma _{\tilde{y}'}^{4}+\sigma _{\tilde{x}'}^{6}+\sigma _{\tilde{y}'}^{6}}{2} \Big )^{\!\!1/6}$.} Furthermore, $\sigma _{\text{R}}^{2}\!=\!\sigma _{\text{R},1}^{2}\!+\!\sigma _{\text{R},2}^{2}$, where $\sigma _{\text{R},1}^{2}$ and $\sigma _{\text{R},2}^{2}$ are defined in (\ref{eq:Rytov}).
\end{corollary}
\begin{proof}
See Appendix \ref{appendix_E}.
\end{proof}
{\color{black}
\vspace{-0.5cm}
\subsection{Two-Decoy-State DV QKD With Finite-Key Effects}
\label{subsect:DVQKD}
Based on the BB84 DV-QKD protocol \cite{BB84}, the decoy-state method utilizes signal states for key transmission, and decoy states for estimating the number of single-photon transmissions \cite{Lo2005}. In our setup, we employ a simple two-decoy-state protocol, using vacuum and weak decoy states, which achieves a key generation rate comparable to protocols with an infinite number of decoy states \cite{Ma2005}. Specifically, the HAP uses a phase-randomized coherent source, encoding bits in the $X$ or $Z$ basis via polarization, as in the standard BB84 scheme. Along with the signal field, vacuum and weak decoy states are generated. Phase randomization ensures the source follows a Poissonian photon-number distribution, where for an average photon number $\mu$, the probability of emitting an $n$-photon pulse is $\exp\!(-\mu)\mu^{n}/n!$. The mean photon numbers are denoted as $\mu_\varrho$, where $\varrho\!\!=\!\!\{\text{s,d,v}\}$ represents signal, weak-decoy, and vacuum states, respectively, with conditions $\mu_{\text{d}}\! \!<\!\! \mu_{\text{s}}\! \!<\!\! 1$ and $\mu_{\text{v}} \!\!=\!\! 0$. At the drone, measurements are performed in randomly chosen $X$ or $Z$ basis. The yield $Y_i$ of an $i$-photon state represents the conditional probability of a detection event given that an $i$-input state is transmitted. The vacuum state estimates the background detection probability $Y_0$, while the weak-decoy state estimates the single-photon yield $Y_1$ and the error rate $e_1$ of the single-photon state. Finally, HAP and the drone conduct key sifting, error correction, and privacy amplification to generate a secure and shared key. 

{\color{black}Conventionally, QKD metrics are statistically estimated assuming an infinite number of key bits transmitted over the channel \cite{Lo2005}. However,} the limited operating duration of the drone restricts the transmission to a finite block of quantum signals, introducing statistical uncertainties in the estimated parameters, commonly referred to as finite-key effects\footnote{{\color{black}By following \cite{Ghalaii2022}, our framework is also applicable for analyzing the security performance of CV-QKD protocols under finite-key effects.}} \cite{Ghalaii2022,Vasylyev2019}. Considering these effects, the QBER is given as \cite{Vasylyev2019}
\begin{align}
\label{eq:QBER}
\text{QBER}\!=\!\frac{\left< E_{\mu _{\text{s}}}Q_{\mu_{\text{s}}}\right>}{\left<Q_{\mu_{\text{s}} }\right>},
\end{align}
where $Q_{\mu_{\text{s}}}$ is the gain of the signal states, $E_{\mu _{\text{s}}}Q_{\mu_{\text{s}}}$ is the overall error gain. Particularly, $\left<\! E_{\mu _{\text{s}}}Q_{\mu_{\text{s}}}\!\right>\!\!=\!\!\int_{0}^{\infty}[e_0Y_0(\tau)\!+\!e_{\text{det}}(1\!-\!\exp(-\mu_{\text{s}}\tau)\!)(1\!-\!Y_0(\tau))]f\!(\tau)\text{d}\tau$, in which $Y_0(\tau)\!\!=\!\!Y_0^{\text{DC}}\!+\!\frac{\frac{1}{2}p_{\text{v}}N\tau}{N(\sum_{\varrho=\text{s,d,v}}\!\exp(-\mu_{\varrho})p_\varrho))}$ with $N$ the total transmitted bits and $p_\varrho$ $(\varrho\!\!=\!\!\text{s,d,v})$ the probability of generating signal, decoy, and vacuum bits, respectively \cite{Vasylyev2019}. Furthermore, $e_{\text{det}}$ is the erroneous detector probability due to the misaligned polarization and stability of the Rx optical system, and $\left<Q_{\mu_{\text{s}} }\right>\!\!=\!\!\int_{0}^{\infty}[1\!-\!\exp(-\mu_{\text{s}}\tau)(1\!-\!Y_0(\tau))]f(\tau)\text{d}\tau$, where $f(\tau)$ is given in (\ref{eq:PDF_tau}). Subsequently, the lower bound of the average SKR, considering finite-key effects, can be expressed as \cite{Vasylyev2019}
\begin{align}
\label{eq:SKR_finite}
\left \langle\mathfrak{R}\right \rangle\!\leq\!\frac{p_{\text{s}}}{2}\!\Big[&\!-\!\left \langle Q_{\mu_{\text{s}}}\right \rangle\! f(\text{QBER})H[\text{QBER}] \nonumber\\
&\!+\!\!\sum_{\iota =X,Z}\!\left \langle Q_{1}^{\iota\text{LB}}\right \rangle\!\!\left(1\!-\!H\!\left[\!\left \langle e_{1}^{\iota\text{UB}}Q_{1}^{\iota\text{LB}}\right \rangle\!/\!\left \langle Q_{1}^{\iota\text{LB}} \right \rangle\!\right] \!\right)\!\!\Big]\!,
\end{align}
where $f(\text{QBER})$ is the bidirectional error correction efficiency, $H[x]\!=\!-x\log_{2}(x)\!-\!(1\!-\!x)\log_{2}(1\!-\!x)$ is the binary Shannon information function, $Q_1$ is the gain of single-photon states, $\text{LB}$ and $\text{UB}$ denote the lower bound and upper bound, respectively. Moreover, $\left \langle Q_{1}^{\iota\text{LB}}\right \rangle\!=\!\int_{0}^{\infty}Y_{1}^{\iota\text{LB}}(\tau)\mu_{\text{s}}\exp(-\mu_{\text{s}})f(\tau)\text{d}\tau$, where $Y_{1}^{\iota\text{LB}}(\tau)$ is given in \cite[(G9)]{Vasylyev2019}. Besides, $e_{1}^{Z\text{UB}}(\tau)\!=\!e_{1}^{X\text{UB}}(\tau)+\theta^{\text{UB}}$ represent the relation between the upper bounds for single-photon error rates in $X$ and $Z$ bases, where $e_{1}^{X\text{UB}}$ is given in \cite[(G10)]{Vasylyev2019} and $\theta^{\text{UB}}$ can be obtained by numerically solving \cite[(G17)]{Vasylyev2019} at a given failure probability $\epsilon_{\text{f}}$. Subsequently, $\left \langle e_{1}^{\iota\text{UB}}Q_{1}^{\iota\text{LB}} \right \rangle\!=\!\int_{0}^{\infty}e_{1}^{\iota\text{UB}}(\tau)Q_{1}^{\iota\text{LB}}(\tau)f(\tau)\text{d}\tau$. Finally, the secret-key length can be calculated by multiplying (\ref{eq:SKR_finite}) with the total transmitted bits $N$, while the drone operating time can be determined by $N/r_\text{N}$ with $r_\text{N}$ the pulse repetition rate.   
}
\vspace{-0.2cm}
\section{Numerical Results and Discussions}
\label{sect:Section5}
\begin{table}[t]
\vspace{-0.2cm}
{\footnotesize
\caption{System and Channel Parameters \cite{Alberto2024,Trinh2022,Andrews2005}}
\vspace{-0.4cm}
\label{Table3}
\begin{center}
\scalebox{1}{
\begin{tabular}{lll}
\hline
\hline
\textbf{System and Channel Parameters}               & \textbf{Notation}            &\textbf{Value}\\
\hline
Optical wavelength  & $\lambda$ & 810 nm\\
Transmission efficiency at zenith  & $\tau_{\text{zen}}$ & 0.78 \\
Atmospheric extinction coefficient  & $\beta_{\text{l}}$ & 0.43 dB/km \\
Optical beam divergence half-angle & $\theta_{\text{div}}$ & 16.5 $\mu$rad\\
Ground turbulence refractive index  & $A$ &$3\!\times\!10^{-13}$ m$^{-2/3}$\\
Ground wind speed  & $v_{\text{g}}$ &5 m/s\\
ORIS's altitude  & $h_{\text{ORIS}}$ &50 m\\
HAP's altitude  & $h_{\text{HAP}}$ &20,000 m\\
LAP's altitude  & $h_{\text{LAP}}$ &300 m\\
ORIS-LAP projected distance  & $d_{\text{LAP}}$ &250 m\\
ORIS-LAP zenith angle  & $\varphi_{\text{r}}$ &$45^{\circ}$\\
{\color{black}ORIS reflectance} &{\color{black}$\tau_{\text{ORIS}}$} &{\color{black}0.9} \\
Receiver aperture radius	& $a$ & 0.045 m\\
Receiver efficiency	& $\tau_{\text{eff}}$ & 50\%\\
\hline
\hline
\textbf{Weak PE Parameters}               & \textbf{Notation}            &\textbf{Value}\\
\hline
Mean hovering in $\tilde{x}'$-axis & $\mu_{\tilde{x}'}$ & 0.3 m\\
Mean hovering in $\tilde{y}'$-axis & $\mu_{\tilde{y}'}$ & 0.2 m\\
Hovering deviation in $\tilde{x}'$-axis & $\sigma_{\tilde{x}'}$ & 0.2 m\\
Hovering deviation in $\tilde{y}'$-axis & $\sigma_{\tilde{y}'}$ & 0.1 m\\
\hline
\hline
\textbf{Moderate PE Parameters}               & \textbf{Notation}            &\textbf{Value}\\
\hline
Mean hovering in $\tilde{x}'$-axis & $\mu_{\tilde{x}'}$ & 0.4 m\\
Mean hovering in $\tilde{y}'$-axis & $\mu_{\tilde{y}'}$ & 0.3 m\\
Hovering deviation in $\tilde{x}'$-axis & $\sigma_{\tilde{x}'}$ & 0.25 m\\
Hovering deviation in $\tilde{y}'$-axis & $\sigma_{\tilde{y}'}$ & 0.2 m\\
\hline
\hline
\textbf{Strong PE Parameters}               & \textbf{Notation}            &\textbf{Value}\\
\hline
Mean hovering in $\tilde{x}'$-axis & $\mu_{\tilde{x}'}$ & 0.5 m\\
Mean hovering in $\tilde{y}'$-axis & $\mu_{\tilde{y}'}$ & 0.4 m\\
Hovering deviation in $\tilde{x}'$-axis & $\sigma_{\tilde{x}'}$ & 0.3 m\\
Hovering deviation in $\tilde{y}'$-axis & $\sigma_{\tilde{y}'}$ & 0.25 m\\
\hline
\end{tabular}}
\end{center}
\vspace{-0.6cm}
}
\end{table}
In this section, we present the analytical results of the average GML and the average PLOB bound of the SKR $\mathfrak{R}$ (bits/use) for different ORIS phase-shift profiles and varying severities of PE, using the main parameters given in Table \ref{Table3}. To confirm the analytical findings, MC simulations are conducted for $10^{7}$ RVs generated for each random parameter. Observe from Table \ref{Table3} that the drone altitude is $h_{\text{LAP}} \!=\! d_{\text{LAP}}\tan\left(\theta_{\text{r}}\right) \!+\! h_{\text{ORIS}} \!=\! 300$ m, and the ORIS-drone distance is $d_2 \!=\! \frac{d_{\text{LAP}}}{\cos\left(\theta_{\text{r}}\right)} \!\cong \!353.55$ m. The HF and EHF principles are valid for intermediate-field ORIS-drone distances if $d_2 \!>\! d_{\text{n}}$, where $d_{\text{n}}$ denotes the minimum intermediate distance defined in \cite[(14)]{Ajam2022}. Upon using the parameters of Table \ref{Table3}, we find that $d_{\text{n}} \!\in\! \left[211.88, 234.94\right]$ m corresponds to $\varphi_{\text{i}} \!\in\! \left[0^{\circ}, 68^{\circ}\right]$. Thus, $d_2$ in our scenario satisfies the condition $d_2 \!>\! d_{\text{n}}$.
\vspace{-0.4cm}
\subsection{Average GML $\tau_{\text{p}}$ With ORIS Phase-Shift Profiles}
\begin{figure*}[t]
\vspace{-0.6cm}
\centering
\subfloat[]{\includegraphics[scale=0.4]{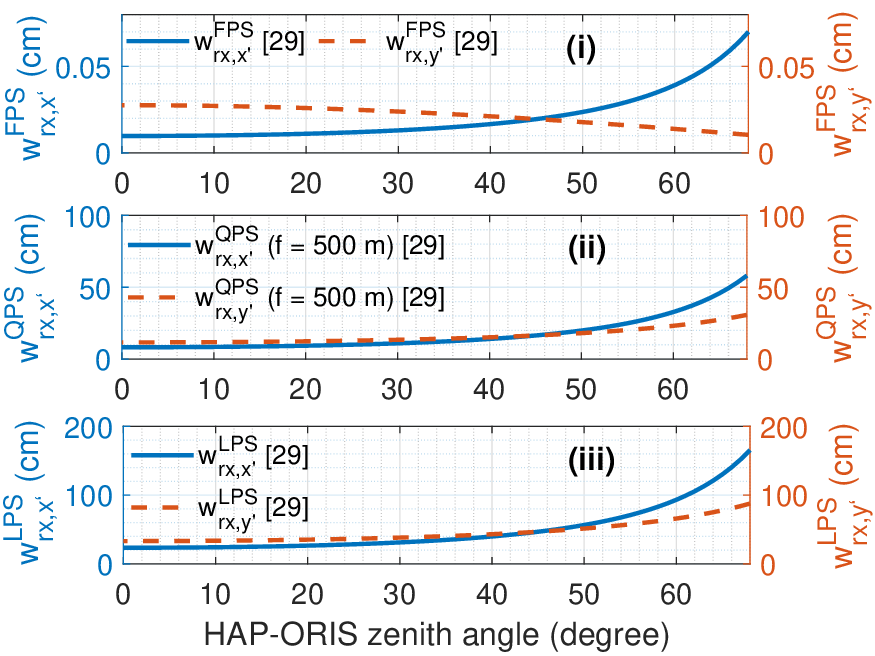}
\label{fig_5a}}
\centering
\subfloat[]{\includegraphics[scale=0.4]{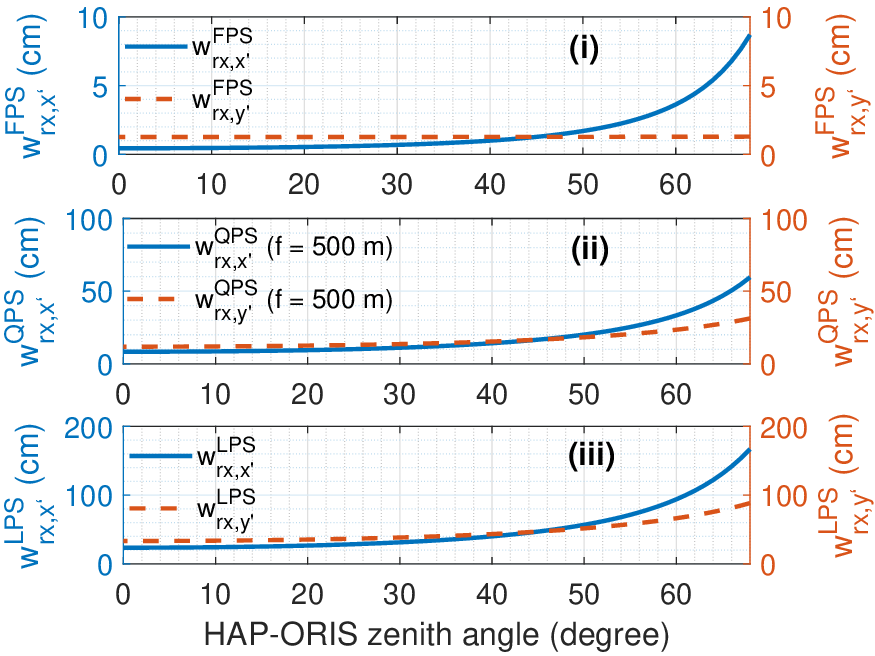}
\label{fig_5b}}
\centering
\subfloat[]{\includegraphics[scale=0.4]{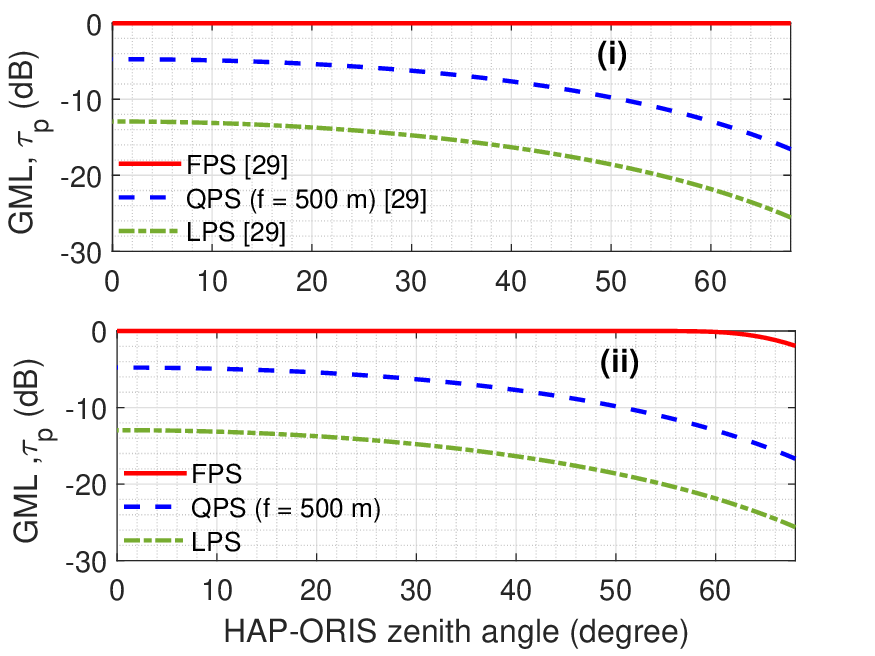}
\label{fig_5c}}
\caption{(a) Rx beam widths $w_{\text{rx},x'}$ and $w_{\text{rx},y'}$ versus zenith angle $\varphi_{\text{i}}$ for FPS (i), QPS (ii) and LPS (iii) profiles \cite{Ajam2024} ; (b) Rx beam widths $w_{\text{rx},x'}$ and $w_{\text{rx},y'}$ versus zenith angle $\varphi_{\text{i}}$ using the framework in this paper for FPS (i), QPS (ii) and LPS (iii) profiles; (c) Comparison of GML $\tau_{\text{p}}$ using the framework in \cite{Ajam2024} (i) versus that in this paper (ii) in the absence of PE.}
\label{fig_5}
\vspace{-0.5cm}
\end{figure*}
\begin{figure}[t]
\centering
\includegraphics[scale=0.5]{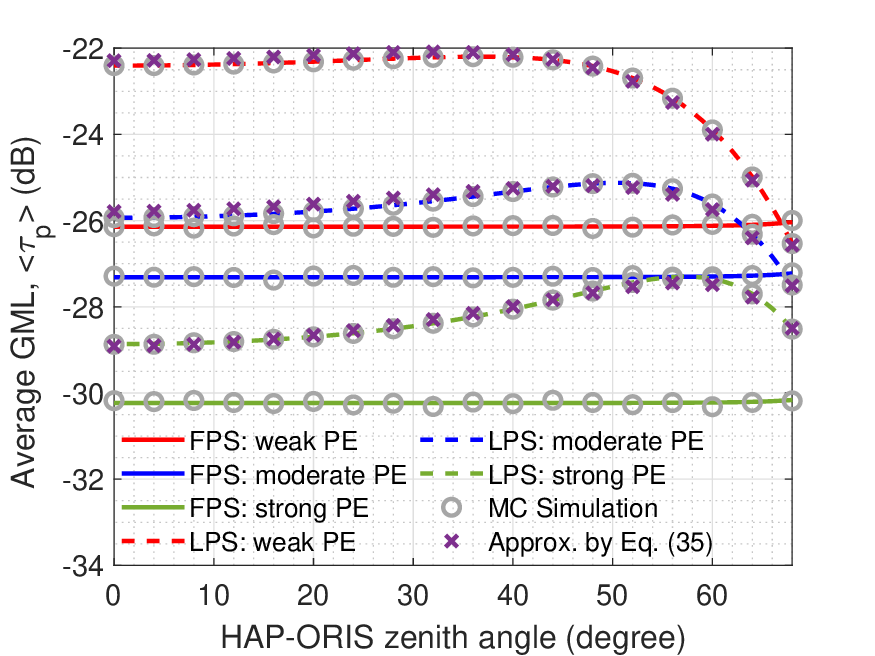}
\caption{Average GML {\color{black}$\left \langle\tau_{\text{p}}\right \rangle$} for FPS and LPS profiles under various PE severities induced by drone hovering fluctuations.}
\label{fig_6}
\vspace{-0.5cm}
\end{figure}
To highlight the atmospheric effects on the optical beam footprint, we compare the Rx beam widths at the drone for different ORIS phase-shift profiles using the frameworks developed in \cite{Ajam2024} and in this paper, as shown in Figs. \ref{fig_5}a and \ref{fig_5}b, respectively. It is observed in Fig. \ref{fig_5}b(i) that atmospheric turbulence-induced beam broadening significantly affects the beam widths of the focused beam induced by the FPS profile. By contrast, this is ignored in Fig. \ref{fig_5}a(i). Specifically, the beam broadening effect is most pronounced at the highest zenith angle of $\varphi_{\text{i}}\!=\!68^\circ$, due to the longest atmospheric path. The broadened beam resulting from turbulence is more than ten times larger than that caused by pure diffraction. On the other hand, for diffractive beams induced by the QPS and LPS profiles, the turbulence-induced beam broadening effect remains insignificant even at $\varphi_{\text{i}}\!=\!68^\circ$. For example, the broadening is only about 1 cm larger than that caused by pure diffraction, as shown in Figs. \ref{fig_5}b(ii) and \ref{fig_5}b(iii) compared to Figs. \ref{fig_5}a(ii) and \ref{fig_5}a(iii). Using the beam width values from Figs. \ref{fig_5}a and \ref{fig_5}b, we analyze the GML $\tau_{\text{p}}$ for all phase-shift profiles without PE based on the frameworks in \cite{Ajam2024} and this paper, as illustrated in Figs. \ref{fig_5}c(i) and \ref{fig_5}c(ii), respectively. As expected, the GML for the FPS profile under turbulence in Fig. \ref{fig_5}c(ii) is not perfectly zero as in Fig. \ref{fig_5}c(i), but it is significantly reduced to $-1.93$ dB at $\varphi_{\text{i}} \!=\! 68^\circ$. Meanwhile, the GML values for the QPS and LPS profiles under turbulence in Fig. \ref{fig_5}c(ii) remain approximately the same as in Fig. \ref{fig_5}c(i), with only about a $0.1$ dB difference at $\varphi_{\text{i}} \!=\! 68^\circ$.

It is evident from Fig. \ref{fig_5}c(ii) that the FPS profile achieves the lowest GML, followed by the QPS and LPS profiles, in the absence of PE. This occurs because the GML is determined by the fraction of power captured by the Rx aperture, with smaller beams resulting in a higher fraction of received power. However, this trend does not hold for the average GML in the presence of PE induced by drone hovering fluctuations, as investigated in Fig. \ref{fig_6} for FPS and LPS profiles. Interestingly, the FPS profile results in a higher average GML than the LPS profile {\color{black}across all zenith angles under strong PE conditions and for most zenith angles, e.g., $\varphi_{\text{i}}\!<\!67^\circ$, in weak-to-moderate PE conditions.} This is because the random fluctuations in the drone position cause higher losses for smaller beam widths. However, at {\color{black}$\varphi_{\text{i}}\!\geq\!67^\circ$ for weak-to-moderate} PE conditions, the average GML of the FPS profile surpasses that of the LPS profile, {\color{black}since} the beam widths induced by the LPS profile become significantly broadened, resulting in higher geometrical loss. Finally, the accuracy of analytical results is validated through MC simulations, showing excellent agreement. {\color{black}The approximated expression in (\ref{eq:GML_approx}) is also validated for the LPS profile, showing a good match with the exact results from (\ref{eq:LP_CL}).}

\begin{figure*}[t]
\vspace{-0.6cm}
\centering
\subfloat[]{\includegraphics[scale=0.4]{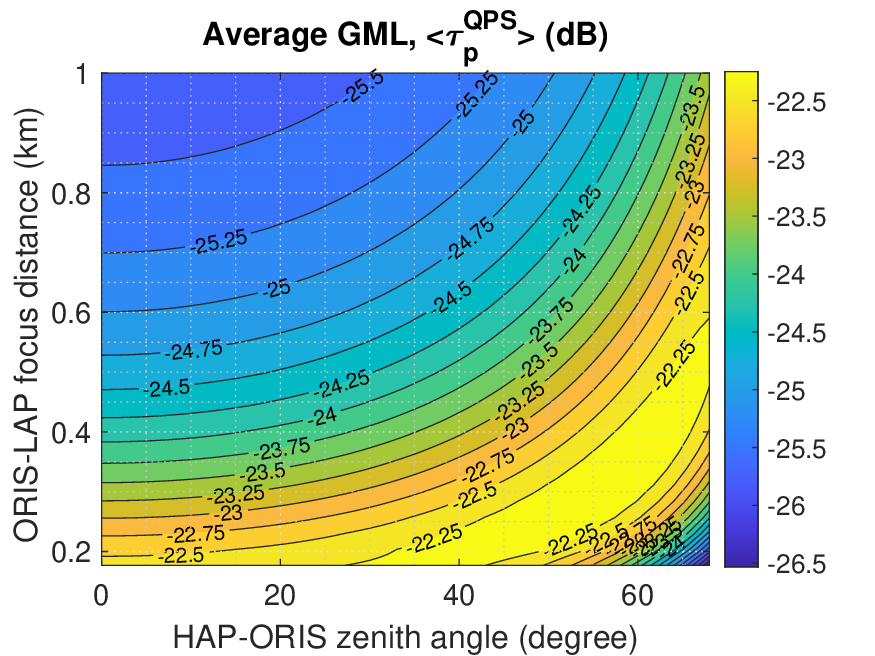}
\label{fig_7a}}
\centering
\subfloat[]{\includegraphics[scale=0.4]{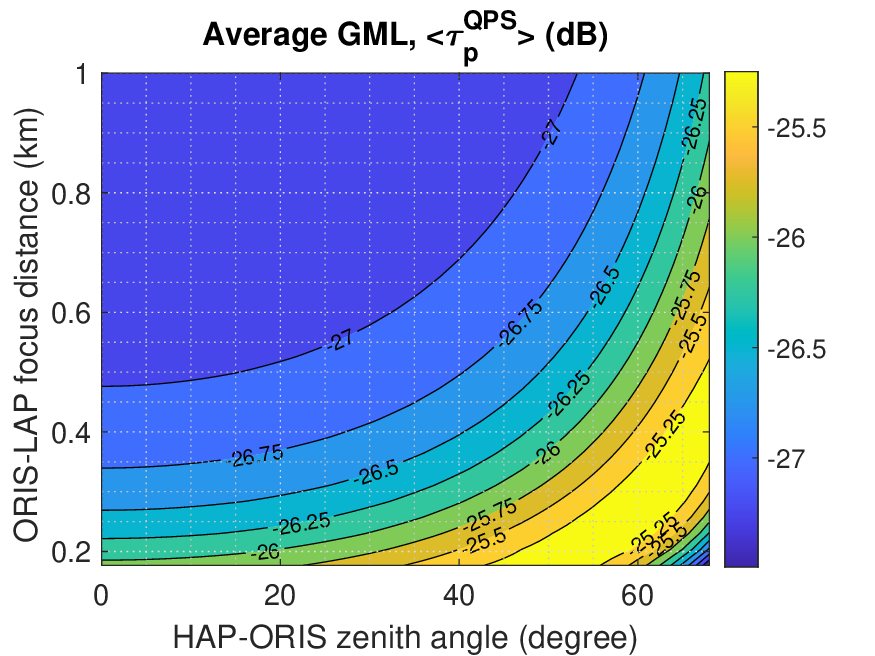}
\label{fig_7b}}
\centering
\subfloat[]{\includegraphics[scale=0.4]{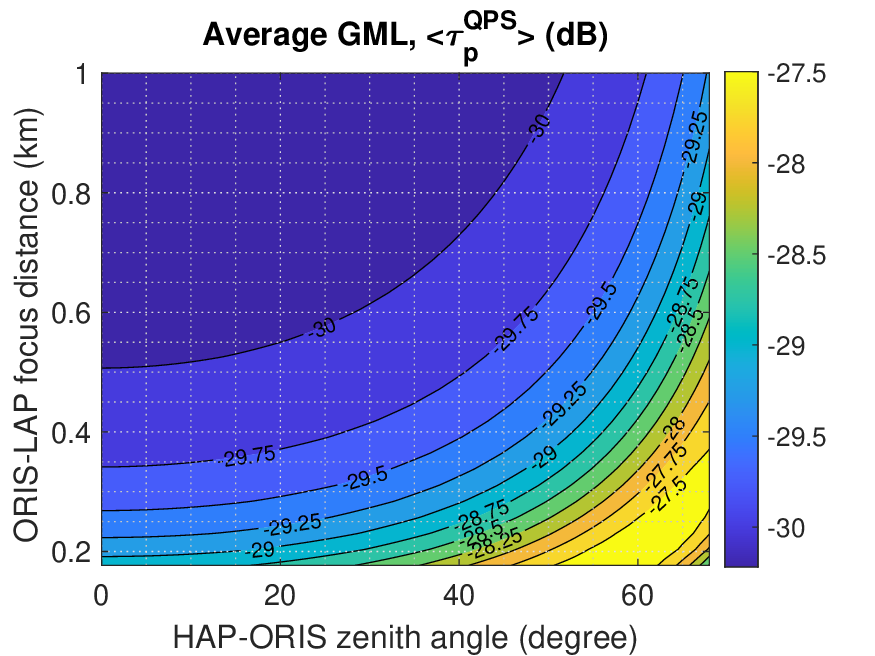}
\label{fig_7c}}
\caption{Average GML {\color{black}$\left \langle\tau_{\text{p}}\right \rangle$} for the QPS profile under various PE severities induced by drone hovering fluctuations. (a) weak PE; (b) moderate PE; (c) strong PE.}
\label{fig_7}
\vspace{-0.5cm}
\end{figure*}

Following Fig. \ref{fig_6}, we continue investigating the impact of PE on the average GML of the QPS profile under weak, moderate, and strong PE conditions in Figs. \ref{fig_7a}, \ref{fig_7b}, and \ref{fig_7c}, respectively. The QPS profile represents an adaptive scheme capable of adjusting the beam widths, encompassing the FPS and LPS as special cases, when the focus distance parameter $f$ is set to infinity and $d_2/2$, respectively. The yellow regions in Figs. \ref{fig_7a}, \ref{fig_7b}, and \ref{fig_7c} highlight the optimal values for the parameter $f$ to achieve the lowest average GML with respect to the zenith angle $\varphi_{\text{i}}$, where the minimum value of $f\!=\!d_2/2\!\cong\!{\color{black}177}$ m corresponds to the special case of using the LPS profile. {\color{black}Apparently}, the QPS profile optimizes the beam width by gradually increasing $f$, thereby narrowing the beam to an optimal size that effectively compensates for drone hovering fluctuations. This optimization is suitable for achieving the lowest {\color{black}possible} GML over {\color{black}low} zenith angles, while maintaining a consistent GML over high zenith angles, e.g., {\color{black}$\varphi_{\text{i}}\!>\!28^\circ$}, $\varphi_{\text{i}}\!>\!{\color{black}43^\circ}$, and $\varphi_{\text{i}}\!>\!{\color{black}51^\circ}$ for {\color{black}weak,} moderate, and strong PE conditions in Figs. {\color{black}\ref{fig_7a},} \ref{fig_7b} and \ref{fig_7c}, respectively.
\vspace{-0.4cm}
\subsection{Average PLOB Bound of The SKR $\mathfrak{R}$}
\begin{figure}[t]
\vspace{-0.1cm}
\centering
\includegraphics[scale=0.5]{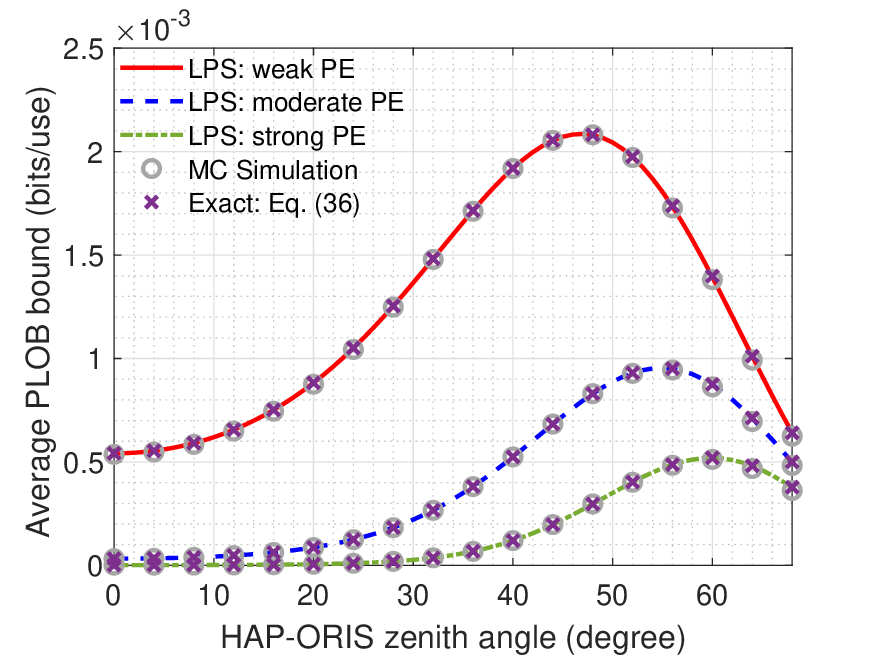}
\caption{Average PLOB bound of the SKR {\color{black}$\left \langle\mathfrak{R}\right \rangle$} (bits/use) {\color{black}for the LPS profile} under various PE severities induced by drone hovering fluctuations. The {\color{black}Gauss-Hermite} polynomial order $G\!=\!{\color{black}100}$.}
\label{fig_8}
\vspace{-0.2cm}
\end{figure}
\begin{figure*}[t]
\vspace{-0.6cm}
\centering
\subfloat[]{\includegraphics[scale=0.4]{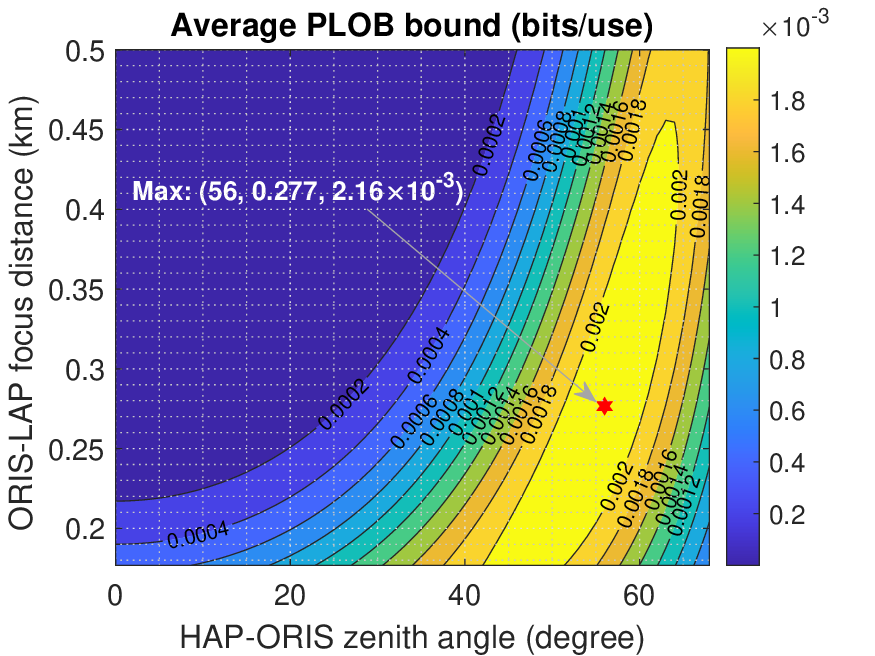}
\label{fig_9a}}
\centering
\subfloat[]{\includegraphics[scale=0.4]{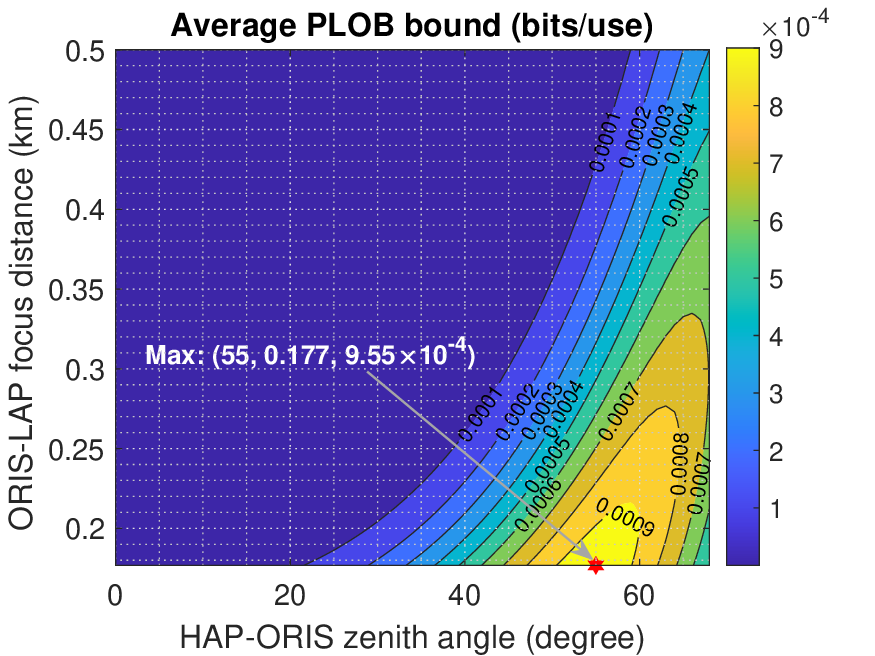}
\label{fig_9b}}
\centering
\subfloat[]{\includegraphics[scale=0.4]{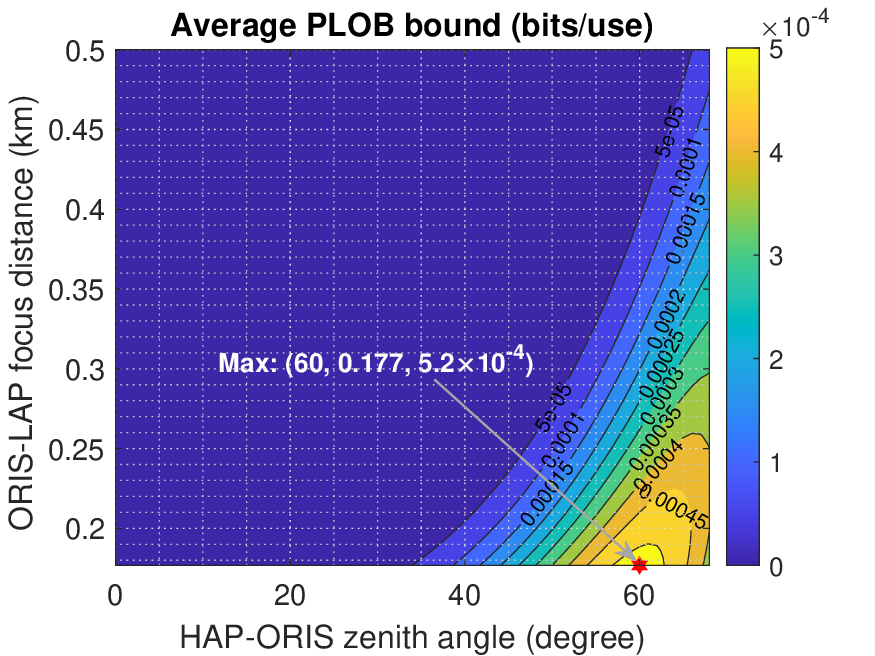}
\label{fig_9c}}
\caption{Average PLOB bound of the SKR {\color{black}$\left \langle\mathfrak{R}\right \rangle$} (bits/use) for the QPS profile under various PE severities induced by drone hovering fluctuations. (a) weak PE; (b) moderate PE; (c) strong PE. The  {\color{black}Gauss-Hermite} polynomial order $G\!=\!{\color{black}100}$.}
\label{fig_9}
\vspace{-0.5cm}
\end{figure*}
In Fig. \ref{fig_8}, we examine the average PLOB bound of the SKR $\mathfrak{R}$ for {\color{black}the LPS profile} versus the PE levels. The analytical PLOB results, derived from {\color{black}Corollary \ref{Corollary2}}, are corroborated by the exact form in (\ref{eq:PLOB_int}) and validated through MC simulations, demonstrating excellent agreement. {\color{black}Fig. \ref{fig_6} previously reveals optimal zenith angles that minimize GML for the LPS profile (e.g., $\varphi_{\text{i}}\!=\!36^\circ$, $50^\circ$, and $57^\circ$ for weak, moderate, and strong PE conditions, respectively). Correspondingly, Fig. \ref{fig_8} also identifies optimal zenith angles for maximizing SKRs under random fluctuations induced by both turbulence and PE (e.g., $\varphi_{\text{i}}\!=\!47^\circ$, $55^\circ$, and $60^\circ$ for weak, moderate, and strong PE conditions, respectively). It is noted that the optimal angles shift to higher values compared to Fig. \ref{fig_6}, resulting in larger beam widths. At a radial distance from the beam centroid, larger beam widths reduce intensity fluctuations caused by turbulence \cite{Andrews2005} and PE \cite{Farid2007,Al2016}, but increase geometrical loss to a finite Rx aperture. This trade-off makes higher zenith angles optimal, balancing fluctuation reduction and geometrical loss.}

Eventually, the PLOB bound of the SKR $\mathfrak{R}$ found for the QPS profile is analyzed under weak, moderate, and strong PE conditions, as shown in Figs. \ref{fig_9a}, \ref{fig_9b}, and \ref{fig_9c}, respectively. The yellow regions in Figs. \ref{fig_9a}, \ref{fig_9b}, and \ref{fig_9c} indicate the optimal values for the parameter $f$ to achieve the highest average SKR relative to the zenith angle $\varphi_{\text{i}}$. Again, the minimum value of $f\!=\!d_2/2\!\cong\!{\color{black}177}$ m corresponds to the special case of using the LPS profile. {\color{black}As seen in Fig. \ref{fig_9a}, the QPS profile is particularly effective under weak PE conditions, where further narrowing the beam width minimizes losses, resulting in the maximum SKR at $\varphi_{\text{i}}\!=\!56^\circ$ with $f\!=\!277$ m. Conversely,} the LPS profile is beneficial for compensating moderate-to-strong PE conditions due to its wider beams, {\color{black}achieving the maximum SKRs at $\varphi_{\text{i}}\!=\! 55^\circ$ and $60^\circ$} in Figs. \ref{fig_9b} and \ref{fig_9c}, respectively. These optimal {\color{black}zenith angles} found for the LPS profiles are {\color{black}consistent with those identified in Fig. \ref{fig_8}.}
{\color{black}
\vspace{-0.3cm}
\subsection{Two-Decoy-State DV QKD With Finite-Key Effects}}
\begin{table}[t]
{\footnotesize
\caption{{\color{black}Two-Decoy-State DV QKD Parameters \cite{Vasylyev2019}}}
\vspace{-0.4cm}
\label{Table4}
\begin{center}
\scalebox{1}{
\begin{tabular}{lll}
\hline
\hline
{\color{black}\textbf{Parameter} }             & {\color{black}\textbf{Notation} }           &{\color{black}\textbf{Value}}\\
\hline
{\color{black}Background yield (dark count)}  & {\color{black}$Y_{0}^{\text{DC}}$} & {\color{black}$5.89\times10^{-7}$} \\
{\color{black}Background error rate}  & {\color{black}$e_{0}$} & {\color{black}50\%} \\
{\color{black}Failure probability}  & {\color{black}$\epsilon_{\text{f}}$} &{\color{black}$10^{-5}$} \\
{\color{black}Mean intensity of signal} & {\color{black}$\mu_{\text{s}}$} & {\color{black}0.8}\\
{\color{black}Mean intensity of weak-decoy state}   & {\color{black}$\mu_{\text{d}}$} &{\color{black}0.1}\\
{\color{black}Pulse repetition rate}  & {\color{black}$r_{\text{N}}$} &{\color{black}200 MHz}\\
{\color{black}Generation prob. of signal bits}  & {\color{black}$p_{\text{s}}$} &{\color{black}65\%}\\
{\color{black}Generation prob. of weak-decoy bits}   & {\color{black}$p_{\text{d}}$} &{\color{black}25\%}\\
{\color{black}Generation prob. of vacuum bits}   & {\color{black}$p_{\text{v}}$} &{\color{black}10\%}\\
{\color{black}Generation prob. of $X$-basis bits}   & {\color{black}$p_{\vartheta}^{X}~(\vartheta=\text{s,d})$} &{\color{black}60\%}\\
{\color{black}Error correction efficiency}  & {\color{black}$f(\text{QBER})$} &{\color{black}1.16}\\
\hline
\end{tabular}}
\end{center}
\vspace{-0.5cm}
}
\end{table}
\begin{figure}[t]
\vspace{-0.1cm}
\centering
\includegraphics[scale=0.5]{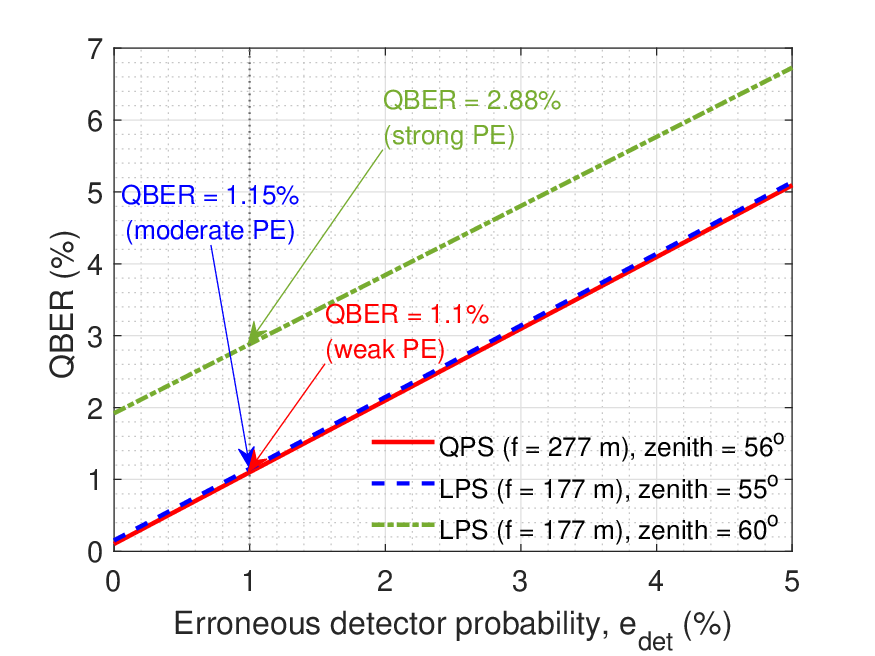}
\caption{{\color{black}QBER versus erroneous detector probability under various settings.}}
\label{fig_10}
\vspace{-0.5cm}
\end{figure}
\begin{figure}[t]
\vspace{-0.1cm}
\centering
\includegraphics[scale=0.5]{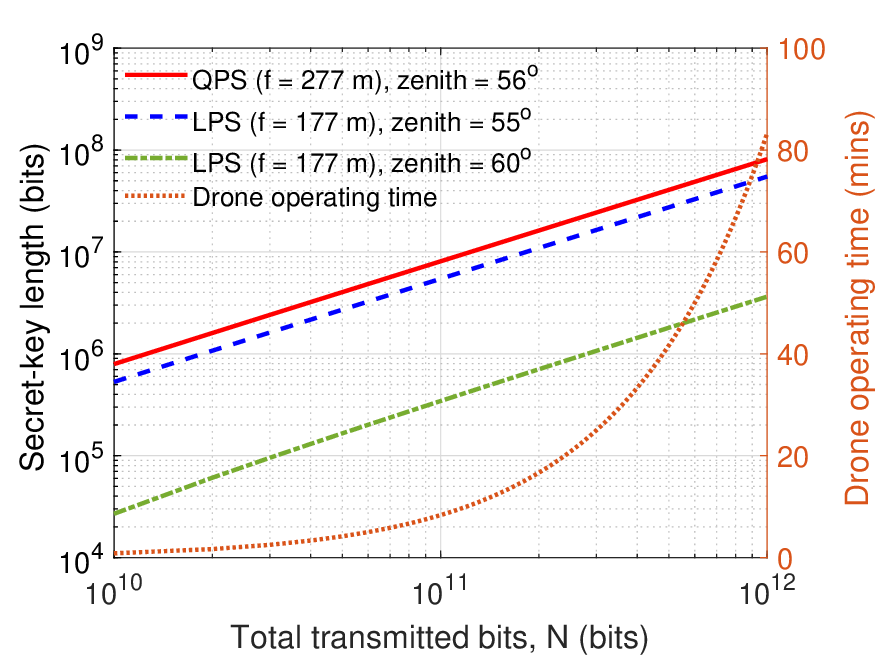}
\caption{{\color{black}Secret-key length versus total transmitted bits within drone operating duration under various settings.}}
\label{fig_11}
\vspace{-0.5cm}
\end{figure}
{\color{black}In this section, we calculate the QBER and secret-key length using the QKD parameters given in Table \ref{Table4}. In Fig. \ref{fig_10}, the QBER in (\ref{eq:QBER}) is numerically investigated as a function of the erroneous detector probability $e_{\text{det}}$, which quantifies error probabilities arising from polarization corrections (as described in Section \ref{subsect:ORIS_pol}) and the stability of the Rx optical system. The investigation is conducted under varying PE levels, considering optimal zenith angles and ORIS phase-shift profile configurations that achieve the maximum PLOB bounds identified in Figs. \ref{fig_9a}, \ref{fig_9b}, and \ref{fig_9c}. It is observed that QBER exhibits a linear dependence on $e_{\text{det}}$, highlighting the critical importance of maintaining an efficient optical receiver in QKD systems, such as ensuring $e_{\text{det}}\!\leq\!1\%$. Consequently, the QBER values corresponding to $e_{\text{det}}\!=\!1\%$ are determined to be $1.1\%$, $1.15\%$, and $2.88\%$ under weak, moderate, and strong PE conditions, respectively.

Using the derived QBER values, the secret-key length as a function of the total transmitted bits during the drone operating duration can be estimated in Fig. \ref{fig_11}, based on the formulation in (\ref{eq:SKR_finite}). The shared secret key length is significantly shorter than the total transmitted bits due to losses, QKD procedures, and finite-key effects. These effects account for statistical fluctuations that reduce the estimated SKR compared to its asymptotic value for infinite data sizes \cite{Lo2005,Ma2005}. This limitation is particularly evident in HAP-to-drone QKD links, where the drone operating time is constrained by battery capacity, typically less than 60 minutes with a 5-kg payload on modern industrial drones \cite{web2024}. Assuming a 60-minute flight duration, the secret-key lengths are approximately $58.58$, $39.53$, and $2.61$ Mbits under weak, moderate, and strong PE conditions, respectively, for $N\!=\!7.2\!\times\!10^{11}$ bits and $r_{\text{N}}\!=\!200$ MHz. It should be noted that longer key lengths allow secure encryption of larger data volumes, adhering to the one-time pad principle, which requires the key length to match or exceed the data size for optimal security \cite{Trinh2024}.}
\section{Conclusions}
\label{sect:Section6}
\vspace{-0.1cm}
The ORIS concept was developed for enhancing QKD links between HAPs and LAPs while mitigating the LAP's hovering fluctuations. By reflecting HAP's incoming beam via a rooftop-mounted ORIS to the terminal beneath the LAP, we established an efficient QKD link. An ORIS facilitates adaptive beam width control through LPS, QPS, and FPS profiles, optimizing the GML at the receiver. This necessitates a robust theoretical framework for accurately characterizing the ORIS-controlled optical beam propagation over atmospheric channels. We employ the EHF principles for the first time to precisely model the atmospheric turbulence effects imposed on ORIS-controlled beams. Our analytical model incorporates the LAP hovering fluctuations, offering a comprehensive framework for ORIS-aided non-terrestrial FSO systems. Utilizing this model, we derive the ultimate PLOB bound for the SKR, {\color{black}and analyze the performance of a two-decoy-state DV-QKD protocol with finite-key effects} over HAP-ORIS-LAP links. Our findings demonstrate that the QPS profile optimizes the SKR at high zenith angles or under mild PE conditions by narrowing the beam to optimal sizes, while the LPS profile is advantageous at low zenith angles {\color{black}or under} the moderate-to-strong PE by diverging the beam to compensate for LAP {\color{black}hovering} fluctuations. These results underscore the efficiency of ORIS in mitigating PEs and optimizing the {\color{black}QKD performance} across diverse conditions.
\appendices
\vspace{-0.2cm}
\section{Proof of Lemma \ref{Lemma1}}
\label{appendix_A}
\vspace{-0.1cm}
Following the framework in \cite{Ajam2024}, with the help of \cite[(54)]{Ajam2024}, the statistical average GML coefficient $\tau_{\text{p}}$ in (\ref{eq:tau_p}) can be approximated by $\tau_{\text{p}}^{\text{LPS}}$, for an LPS profile considering drone hovering fluctuations, written as
\begin{align}
\label{eq:tau_p_LP}
 {\color{black}\left \langle \!\tau_{\text{p}}^{\text{LPS}}\right \rangle}\!\!=&C_{\text{LPS}}\!\!\iint_{\!-\frac{a\sqrt{\pi}}{2}}^{\frac{a\sqrt{\pi}}{2}}\!\!\iint_{\!-\infty}^{\infty}\!\!\exp\!\!\left (\!\! -\frac{k^2\!\sin^{2}\!\!\left ( \theta _{\text{r}} \right )\!\left (x'\!\!+\!\!\tilde{x}'\right )^{\!2}\!\mathcal{R}\!\left \{ \!b_{x'\!,\text{LPS}} \!\right \}}{2d_2^2\left | b_{x'\!,\text{LPS}} \right |^{2}} \!\!\right )\nonumber\\
&\!\!\!\!\!\!\times\!\exp\!\!\left (\!\! -\frac{k^2\!\left (y'\!+\!\tilde{y}'\right )^{\!2}\!\mathcal{R}\!\left \{ \!b_{y'\!,\text{LPS}} \!\right \}}{2d_2^2\left | b_{y'\!,\text{LPS}} \right |^{2}} \!\!\right )\!\!f_{\tilde{x}'}\!\!\left ( \tilde{x}' \right )\!\!f_{\tilde{y}'}\!\!\left ( \tilde{y}' \right )\!\text{d}x'\!\text{d}y'\!\text{d}\tilde{x}'\!\text{d}\tilde{y}'\!,
\end{align}
where $x',y'\!\in\!\left[-\frac{a\sqrt{\pi}}{2},\frac{a\sqrt{\pi}}{2}\right]$ and $\tilde{x}',\tilde{y}'\!\in\!\left(-\infty,\infty\right)$. Furthermore, we have $C_{\text{LPS}}\!=\!\frac{2P_{\text{t}}\sin\left ( \theta _{\text{i}} \right )\!\sin\left ( \theta _{\text{r}} \right )\pi}{\lambda ^{2}w^{2}\!\left ( d_1 \right )d_2^2\left | b_{x',\text{LPS}} \right |\left | b_{y',\text{LPS}} \right |}$ with $b_{x',\text{LPS}}\!=\!\frac{\sin^{2}\left ( \theta _{\text{i}} \right )}{w^{2}\!\left ( d_1 \right )}\!+\!\frac{jk\sin^{2}\left ( \theta _{\text{i}} \right )}{2R\left ( d_1 \right )}\!+\!\frac{jk\sin^{2}\left ( \theta _{\text{r}} \right )}{2d_2}$ and $b_{y',\text{LPS}}\!=\!\frac{1}{w^{2}\!\left ( d_1 \right )}\!+\!\frac{jk}{2R\left ( d_1 \right )}\!+\!\frac{jk}{2d_2}$. $f_{\tilde{x}'}\!\!\left ( \tilde{x}' \right )$ and $f_{\tilde{y}'}\!\!\left ( \tilde{y}' \right )$ are the Gaussian probability density functions of the hovering fluctuations $\tilde{x}'$ and $\tilde{y}'$, respectively, given by
$f_{\nu}\!\!\left ( \nu \right )\!=\!\frac{1}{\sqrt{2\pi\sigma _{\nu}^{2}}}\exp\!\left (- \frac{\left (\nu-\mu _{\nu}  \right )^{2}}{2\sigma _{\nu}^{2}} \right ),~\nu =\left \{ \tilde{x}',\tilde{y}' \right \}$.
Using \cite[(4.3.13)]{Ng1969} along with a change of variables, we arrive at the following result
\begin{align}
\label{eq:integral}
\int_{-\infty }^{\infty }\!\!\text{erf}\!\left ( \alpha\nu\!+\!\beta \right )\!\!\frac{\exp\!\!\left (\!- \frac{\left (\nu-\mu _{\nu} \right )^{2}}{2\sigma _{\nu}^{2}}\! \right )}{\sqrt{2\pi\sigma _{\nu}^{2}}}\text{d}\nu\!=\!\text{erf}\!\left (\!\! \frac{\alpha\mu _{\nu} \!+\! \beta}{\sqrt{1\!+\!2\alpha^{2}\sigma _{\nu}^{2}}} \!\right )\!\!.
\end{align}
By invoking (\ref{eq:integral}) and \cite[(2.33.1)]{Grad2015}, (\ref{eq:tau_p_LP}) can be solved and a closed-form expression is obtained in (\ref{eq:LP_CL}), where $w_{\text{rx},x'}^{\text{LPS}}\!\!=\!\!w\!\left ( d_1\right )\!\frac{\left | \sin\left ( \theta_{\text{r}}  \right )\right|}{\left | \sin\left ( \theta_{\text{i}}  \right )\right| } \sqrt{\!\varepsilon\!\left (\! \frac{\sin^{2}\left ( \theta_{\text{i}}  \right )}{\sin^{2}\left ( \theta_{\text{r}}  \right )}\Lambda _{1}  \!\right )^{\!2}\!+\!\left (\! \frac{\sin^{2}\left ( \theta_{\text{i}}  \right )}{\sin^{2}\left ( \theta_{\text{r}}  \right )}\Lambda _{2} \!+\!1 \!\right )^{\!2}}$ and $w_{\text{rx},y'}^{\text{LPS}}\!=\!w\!\left ( d_1\right ) \!\sqrt{\varepsilon\Lambda _{1}^{2}\!+\!\left (\Lambda _{2}\!+\!1\right )^{2}}$ are the equivalent beam widths induced by the LPS profile at the Rx aperture in the $x'$ and $y'$ axes, respectively. Furthermore, $\Lambda _{1} \!=\!\frac{2d_2}{kw^2(d_1)}$ and $\Lambda _{2} \!=\!\frac{d_2}{R(d_1)}$ characterize the diffraction and refraction effects, respectively. Since $d_1\gg z_{\text{R1}}$, we have $R(d_1)\!=\!d_1\!\left[1\!+\!\frac{z_{\text{R1}}}{d_1} \right]\!\approx\! d_1$, thus $\Lambda_{2} \!=\!\frac{d_2}{d_1}$. As $d_1\gg d_2$, $\Lambda _{2}\rightarrow0$ and can be omitted, which gives the results of $w_{\text{rx},x'}^{\text{LPS}}$ and $w_{\text{rx},y'}^{\text{LPS}}$ in Lemma \ref{Lemma1}. This completes the proof.
\vspace{-0.2cm}
\section{Proof of Lemma \ref{Lemma2}}
\label{appendix_B}
\vspace{-0.1cm}
Following the framework in \cite{Ajam2024}, with the help of \cite[(55)]{Ajam2024}, the statistical average GML coefficient $\tau_{\text{p}}$ in (\ref{eq:tau_p}) derived for a QPS profile by considering the drone hovering fluctuations can be approximated by $\tau_{\text{p}}^{\text{QPS}}$ as
\begin{align}
\label{eq:tau_p_QP}
 {\color{black}\left \langle \!\tau_{\text{p}}^{\text{QPS}}\right \rangle}\!\!=&C_{\text{QPS}}\!\!\iint_{\!-\frac{a\sqrt{\pi}}{2}}^{\frac{a\sqrt{\pi}}{2}}\!\!\iint_{\!-\infty}^{\infty}\!\!\exp\!\!\left (\!\! -\frac{k^2\!\sin^{2}\!\!\left ( \theta _{\text{r}} \right )\!\left (\!x'\!\!+\!\!\tilde{x}'\right )^{\!2}\!\mathcal{R}\!\left \{ \!b_{x'\!,\text{QPS}} \!\right \}}{2d_2^2\left | b_{x'\!,\text{QPS}} \right |^{2}} \!\!\right )\nonumber\\
&\!\!\!\!\!\!\!\times\!\exp\!\!\left (\!\! -\frac{k^2\!\left (y'\!+\!\tilde{y}'\right )^{\!2}\!\mathcal{R}\!\left \{ \!b_{y'\!,\text{QPS}} \!\right \}}{2d_2^2\left | b_{y'\!,\text{QPS}} \right |^{2}} \!\!\right )\!\!f_{\tilde{x}'}\!\!\left ( \tilde{x}' \right )\!\!f_{\tilde{y}'}\!\!\left ( \tilde{y}' \right )\!\text{d}x'\!\text{d}y'\!\text{d}\tilde{x}'\!\text{d}\tilde{y}'\!,
\end{align}
where we have $C_{\text{QPS}}\!=\!\frac{2P_{\text{t}}\sin\left ( \theta _{\text{i}} \right )\!\sin\left ( \theta _{\text{r}} \right )\pi}{\lambda ^{2}w^{2}\!\left ( d_1 \right )d_2^2\left | b_{x',\text{QPS}} \right |\left | b_{y',\text{QPS}} \right |}$ with $b_{x',\text{QPS}}\!=\!\frac{\sin^{2}\left ( \theta _{\text{i}} \right )}{w^{2}\!\left ( d_1 \right )}\!+\!\frac{jk\sin^{2}\left ( \theta _{\text{r}} \right )}{4f}$ and $b_{y',\text{QPS}}\!=\!\frac{1}{w^{2}\!\left ( d_1 \right )}\!+\!\frac{jk}{4f}$. Similar to Appendix \ref{appendix_A}, by invoking (\ref{eq:integral}) and \cite[(2.33.1)]{Grad2015}, (\ref{eq:tau_p_QP}) can be solved and a closed-form expression is obtained in (\ref{eq:QP_CL}). This completes the proof.
\vspace{-0.3cm}
\section{Proof of Lemma \ref{Lemma3}}
\label{appendix_C}
\vspace{-0.1cm}
Following the framework in \cite{Ajam2024}, with the help of \cite[(56)]{Ajam2024}, the statistical average GML coefficient $\tau_{\text{p}}$ in (\ref{eq:tau_p}) can be approximated by $\tau_{\text{p}}^{\text{FPS}}$ for an FPS profile considering drone hovering fluctuations as
\begin{align}
\label{eq:tau_p_FP}
 {\color{black}\left\langle \!\tau_{\text{p}}^{\text{FPS}}\right \rangle}\!\!=\!C_{\text{FPS}}\!\!\iint_{-\frac{a\sqrt{\pi}}{2}}^{\frac{a\sqrt{\pi}}{2}}\!\!\iint_{-\infty}^{\infty}&\!\exp\!\!\Bigg [\! -\frac{2\!\left (x'\!+\!\tilde{x}'\right )^{\!2}}{(\!w_{\text{rx},x'}^{\text{FPS}}\!)^{\!2}}-\frac{2\!\left (y'\!+\!\tilde{y}'\right )^{\!2}}{(\!w_{\text{rx},y'}^{\text{FPS}}\!)^{\!2}} \!\Bigg ]\nonumber\\
&\!\times\!f_{\tilde{x}'}\!\!\left ( \tilde{x}' \right )\!f_{\tilde{y}'}\!\!\left ( \tilde{y}' \right )\!\text{d}x'\!\text{d}y'\!\text{d}\tilde{x}'\!\text{d}\tilde{y}'\!,
\end{align}
where $C_{\text{FPS}}\!=\!\frac{2P_{\text{t}}\pi w^{2}\!\left ( d_1 \right )\sin\left ( \theta _{\text{r}} \right )}{\lambda ^{2}d_2^2\sin\left ( \theta _{\text{i}} \right )}$, $w_{\text{rx},x'}^{\text{FPS}}\!=\!w\!\left ( d_1\right )\!\frac{\left|\sin\left ( \theta_{\text{i}}  \right )\right|}{\left|\sin\left ( \theta_{\text{r}}  \right )\right|}\sqrt{\varepsilon\Lambda_1^2}$ and $w_{\text{rx},y'}^{\text{FPS}}\!=\!w\!\left ( d_1\right )\!\sqrt{\varepsilon\Lambda_1^2}$ are the equivalent beam widths induced by the FPS profile at the Rx aperture in the $x'$ and $y'$ axes, respectively. Similar to Appendix \ref{appendix_A}, by invoking (\ref{eq:integral}) and \cite[(2.33.1)]{Grad2015}, (\ref{eq:tau_p_FP}) can be solved and a closed-form expression is obtained in (\ref{eq:FP_CL}). This completes the proof.
\vspace{-0.3cm}
{\color{black}
\section{Proof of Corollary \ref{Corollary1}}
\label{appendix_D}
\vspace{-0.1cm}
Following the theoretical framework in \cite[Appendix]{Farid2007}, the instantaneous $\tau_{\text{p}}^{\text{LPS}}$ and $\tau_{\text{p}}^{\text{QPS}}$ extracted from (\ref{eq:tau_p_LP}) and (\ref{eq:tau_p_QP}), respectively, can be approximated by (\ref{eq:tau_p_U}). Since $\tilde{x}'$ and $\tilde{y}'$, are i.n.i.d. Gaussian RVs, i.e., $\tilde{x}'\!\!\sim\!\! \mathcal{N}(\mu_{\tilde{x}'},\sigma_{\tilde{x}'}^2)$ and $\tilde{y}'\!\!\sim\!\! \mathcal{N}(\mu_{\tilde{y}'},\sigma_{\tilde{y}'}^2)$, the radial displacement due to drone hovering follows the Beckmann distribution with the probability density function given in \cite[(17)]{Al2016}. Following the derivation steps in \cite[(18), (19), and (20)]{Al2016}, $\left\langle \tau_{\text{p}}^{\text{U}}\right \rangle$ with $\text{U}\!\!\in\!\!\{\text{LPS},\text{QPS}\}$ can be derived as (\ref{eq:GML_approx}). This completes the proof.
}
\vspace{-0.3cm}
\section{Proof of Corollary \ref{Corollary2}}
\label{appendix_E}
\vspace{-0.1cm}
We have $I_{\text{a}}\!=\!I_{\text{a,1}}I_{\text{a,2}}$, where $I_{\text{a,1}}$ and $I_{\text{a,2}}$ are independent log-normal RVs, due to the distinct atmospheric paths of the HAP-ORIS and ORIS-drone links, respectively. Since $I_{\text{a,1}}\!\sim\! \mathcal{LN}\!(-\frac{\sigma _{\text{R},1}^{2}}{2},\sigma _{\text{R},1}^{2})$ and $I_{\text{a,2}}\!\sim\! \mathcal{LN}\!(-\frac{\sigma _{\text{R},2}^{2}}{2},\sigma _{\text{R},2}^{2})$, it is straightforward to obtain that $I_{\text{a}}\!\sim\!\mathcal{LN}(-\frac{\sigma _{\text{R},1}^{2}}{2}\!-\!\frac{\sigma _{\text{R},2}^{2}}{2},\sigma _{\text{R},1}^{2}\!+\!\sigma _{\text{R},2}^{2})$, where $\sigma _{\text{R},1}^{2}$ and $\sigma _{\text{R},2}^{2}$ are defined in (\ref{eq:Rytov}). As a result, $f\!\left ( I_{\text{a}} \right )$ {\color{black}can be} expressed as
\begin{align}
\label{eq:LN_Ia}
f\!\left ( I_{\text{a}} \right )\!=\!\frac{1}{I_{\text{a}}\sqrt{2\pi\sigma _{\text{R}}^{2}}}\exp\!\!\Bigg [ -\frac{\left ( \ln\left (  I_{\text{a}} \right ) \!+\!\frac{\sigma _{\text{R}}^{2}}{2}\right )^{\!2}}{2\sigma _{\text{R}}^{2}} \Bigg ],
\end{align}
where $\sigma _{\text{R}}^{2}\!=\!\sigma _{\text{R},1}^{2}\!+\!\sigma _{\text{R},2}^{2}$. Additionally, as {\color{black}$\tau\!=\!\tau _{\text{eff}}\tau _{\text{ORIS}}\tau _{\text{l}}I_{\text{a}}\tau_{\text{p}}$} is truncated at 1 to preserve the canonical commutation of the input-output quantum relationship as seen in (\ref{eq:PLOB_int}), $I_{\text{a}}$ is restricted to $\left[0, 1/\tau _{\text{eff}}{\color{black}\tau _{\text{ORIS}}}\tau _{\text{l}}\tau_{\text{p}}\right]$. However, due to the small values of $\tau _{\text{eff}}{\color{black}\tau _{\text{ORIS}}}\tau _{\text{l}}\tau_{\text{p}}$, we can assume that $I_{\text{a}}\in\left[0,\infty\right)$, while satisfying the canonical commutation relationship via $\mathbb{E}\left[ I_{\text{a}} \right]\!=\!1$ \cite{Trinh2022}. {\color{black}Considering that $I_{\text{a}}$ and $\tau_{\text{p}}$ are independent RVs, the probability distribution of $\tau$ can be expressed as $f\!(\tau)\!=\!\int\! f\!(\tau|I_{\text{a}})f\!\left ( I_{\text{a}} \right )\text{d}I_{\text{a}}$, where $f\!(\tau|I_{\text{a}})\!=\!\frac{1}{\tau _{\text{eff}}\tau _{\text{ORIS}}\tau _{\text{l}}I_{\text{a}}}f_{\tau_{\text{p}}}\!\left(\!\frac{\tau}{\tau _{\text{eff}}\tau _{\text{ORIS}}\tau _{\text{l}}I_{\text{a}}}\!\right)$ is the conditional probability given a turbulence state $I_{\text{a}}$ with $f_{\tau_{\text{p}}}\!\left(\cdot\right)$ the probability distribution function of $\tau_{\text{p}}$. Since the radial displacement due to drone hovering follows the Beckmann distribution in \cite[(17)]{Al2016}, $f_{\tau_{\text{p}}}\!\left(\cdot\right)$ can be derived in a closed-form approximation as 
%
$f_{\tau_{\text{p}}}\!\left(\tau_{\text{p}}\right)\!\approx\!\frac{\gamma_{\text{mod}}^{2}}{\left(A_{\text{mod}}\right)^{\!\gamma_{\text{mod}}^{2}}}\tau_{\text{p}}^{\gamma_{\text{mod}}^{2}\!-\!1}$, $(0\!\leq\!\tau_{\text{p}}\!\leq\!A_{\text{mod}})$ \cite[(11)]{Ruben2016},
%
where $A_{\text{mod}}$ and $\gamma_{\text{mod}}$ are given in Corollary \ref{Corollary2}. Following derivation steps in \cite[(13) and (14)]{Farid2007} and with the help of (\ref{eq:LN_Ia}), $f(\tau)$, $\tau\in\left[0,\infty\right)$, can be derived as
\begin{align}
\label{eq:PDF_tau}
f\!\!\left (\tau \right )\!\!=&\frac{\gamma_\text{mod}^{2}}{2\!\left ( A_\text{mod} \tau_{\text{eff}}\tau _{\text{ORIS}}\tau _{\text{l}}\right )^{\gamma_\text{mod}^{2}}}\tau^{\gamma_\text{mod} ^{2}-1}\! \nonumber\\
&\!\!\!\!\!\!\times\!\!\text{erfc}\!\!\left (\! \frac{\text{ln}\!\!\left (\!\frac{\tau}{A_\text{mod}\tau _{\text{eff}}\tau _{\text{ORIS}}\tau _{\text{l}}}\! \right )\!\!+\!\!\Upsilon }{\sqrt{2}\sigma _{\text{R}}} \!\right )\! \!\exp\!\!\left (\!\frac{\sigma _{\text{R}}^{2}}{2}\gamma_\text{mod}^{2}\!\left (\! 1\!+\!\gamma_\text{mod} ^{2} \right )\!\!\right )\!\!,
\end{align}
where $\Upsilon\!=\!\frac{\sigma _{\text{R}}^{2}}{2}\left ( 1+2\gamma_\text{mod} ^{2} \right )$. Substituting (\ref{eq:PDF_tau}) into (\ref{eq:PLOB_int}), making a change of variables, and applying the Gauss-Hermite polynomial $\int_{-\infty }^{\infty }f\!\left ( x \right )\textnormal{d}x\!\approx\! \sum_{g=1}^{G}\!w_g\!\exp\!\left ( x_g^{2} \right )\!f\!\left ( x_{g} \right )$ \cite[Table 25.10]{Abra1972}, (\ref{eq:PLOB_int}) can be derived as a closed-form expression in (\ref{eq:PLOB_CL}) for the LPS and QPS profiles.} This completes the proof.

%
%
%
%
%
%
%
%
%

\end{document}